\documentclass[prd,superscriptaddress,nofootinbib]{revtex4}
\usepackage{amsmath,amssymb,epsfig}
\usepackage[hypertex]{hyperref}
\usepackage{natbib,ifthen}

\begin{document}

\newcommand{\fixme}[1]{{\textbf{Fixme: #1}}}
\newcommand{\detD}{{\det\!\cld}}
\newcommand{\clh}{\mathcal{H}}
\newcommand{\ud}{{\rm d}}
\renewcommand{\eprint}[1]{\href{http://arxiv.org/abs/#1}{#1}}
\newcommand{\adsurl}[1]{\href{#1}{ADS}}
\newcommand{\ISBN}[1]{\href{http://cosmologist.info/ISBN/#1}{ISBN: #1}}
\newcommand{\vort}{\varpi}
\newcommand\ba{\begin{eqnarray}}
\newcommand\ea{\end{eqnarray}}
\newcommand\be{\begin{equation}}
\newcommand\ee{\end{equation}}
\newcommand\lagrange{{\cal L}}
\newcommand\cll{{\cal L}}
\newcommand\cln{{\cal N}}
\newcommand\clx{{\cal X}}
\newcommand\clz{{\cal Z}}
\newcommand\clv{{\cal V}}
\newcommand\cld{{\cal D}}
\newcommand\clt{{\cal T}}

\newcommand\clo{{\cal O}}
\newcommand{\cla}{{\cal A}}
\newcommand{\clp}{{\cal P}}
\newcommand{\clr}{{\cal R}}
\newcommand{\uD}{{\mathrm{D}}}
\newcommand{\calE}{{\cal E}}
\newcommand{\calB}{{\cal B}}
\newcommand{\curl}{\,\mbox{curl}\,}
\newcommand\del{\nabla}
\newcommand\Tr{{\rm Tr}}
\newcommand\half{{\frac{1}{2}}}
\newcommand\fourth{{1\over 8}}
\newcommand\bibi{\bibitem}
\newcommand{\kf}{\beta}
\newcommand{\kfprod}{\alpha}
\newcommand\calS{{\cal S}}
\renewcommand\H{{\cal H}}
\newcommand\K{{\rm K}}
\newcommand\mK{{\rm mK}}
\newcommand\synch{\text{syn}}
\newcommand\opacity{\tau_c^{-1}}

\newcommand{\Psil}{\Psi_l}
\newcommand{\bsigma}{{\bar{\sigma}}}
\newcommand{\bI}{\bar{I}}
\newcommand{\bq}{\bar{q}}
\newcommand{\bv}{\bar{v}}
\renewcommand\P{{\cal P}}
\newcommand{\numfrac}[2]{{\textstyle \frac{#1}{#2}}}

\newcommand{\la}{\langle}
\newcommand{\ra}{\rangle}

\newcommand{\Omtot}{\Omega_{\mathrm{tot}}}
\newcommand\xx{\mbox{\boldmath $x$}}
\newcommand{\phpr} {\phi'}
\newcommand{\gam}{\gamma_{ij}}
\newcommand{\sqgam}{\sqrt{\gamma}}
\newcommand{\delk}{\Delta+3{\K}}
\newcommand{\dph}{\delta\phi}
\newcommand{\om} {\Omega}
\newcommand{\dom}{\delta^{(3)}\left(\Omega\right)}
\newcommand{\rar}{\rightarrow}
\newcommand{\Rar}{\Rightarrow}
\newcommand\gsim{ \lower .75ex \hbox{$\sim$} \llap{\raise .27ex \hbox{$>$}} }
\newcommand\lsim{ \lower .75ex \hbox{$\sim$} \llap{\raise .27ex \hbox{$<$}} }
\newcommand\bigdot[1] {\stackrel{\mbox{{\huge .}}}{#1}}
\newcommand\bigddot[1] {\stackrel{\mbox{{\huge ..}}}{#1}}
\newcommand{\Mpc}{\text{Mpc}}
\newcommand{\Al}{{A_l}}
\newcommand{\Bl}{{B_l}}
\newcommand{\eAl}{e^\Al}
\newcommand{\ix}{{(i)}}
\newcommand{\ixp}{{(i+1)}}
\renewcommand{\k}{\beta}
\newcommand{\HD}{\mathrm{D}}

\newcommand{\nonflat}[1]{#1}
\newcommand{\Cgl}{C_{\text{gl}}}
\newcommand{\Cgltwo}{C_{\text{gl},2}}
\newcommand{\He}{{\rm{He}}}
\newcommand{\Mhz}{{\rm MHz}}
\newcommand{\vx}{{\mathbf{x}}}
\newcommand{\ve}{{\mathbf{e}}}
\newcommand{\vv}{{\mathbf{v}}}
\newcommand{\vk}{{\mathbf{k}}}
\newcommand{\vn}{{\mathbf{n}}}

\newcommand{\vnhat}{{\hat{\mathbf{n}}}}
\newcommand{\vkhat}{{\hat{\mathbf{k}}}}
\newcommand{\taueps}{{\tau_\epsilon}}

\newcommand{\vgrad}{{\mathbf{\nabla}}}
\newcommand{\fbarln}{\bar{f}_{,\ln\epsilon}(\epsilon)}


\title{The linear power spectrum of observed source number counts}

\author{Anthony Challinor}
\affiliation{Institute of Astronomy and Kavli Institute for Cosmology, Madingley Road, Cambridge, CB3 0HA, U.K.}
 \affiliation{DAMTP, Centre
for Mathematical Sciences, Wilberforce Road, Cambridge CB3 0WA, U.K.}

\author{Antony Lewis}
\homepage{http://cosmologist.info}
\affiliation{Department of Physics \& Astronomy, University of Sussex, Brighton BN1 9QH, U.K.}
\affiliation{Institute of Astronomy and Kavli Institute for Cosmology, Madingley Road, Cambridge, CB3 0HA, U.K.}



\begin{abstract}
We relate the observable number of sources per solid angle and redshift to the underlying proper source density and velocity, background evolution and line-of-sight potentials. We give an exact result in the case of linearized perturbations assuming general relativity. This consistently includes contributions of the source density perturbations and redshift distortions, magnification, radial displacement, and various additional linear terms that are small on sub-horizon scales. In addition we calculate the effect on observed luminosities, and hence the result for sources observed as a function of flux, including magnification bias and radial-displacement effects. We give the corresponding linear result for a magnitude-limited survey at low redshift, and discuss the angular power spectrum of the total count distribution. We also calculate the cross-correlation with the CMB polarization and temperature including Doppler source terms, magnification, redshift distortions and other velocity effects for the sources, and discuss why the contribution of redshift distortions is generally small. Finally we relate the result for source number counts to that for the brightness of line radiation, for example 21-cm radiation, from the sources.
\end{abstract}

\maketitle


\section{Introduction}

Redshift surveys of galaxies and other sources consist of a set of source positions along with measured redshifts and other information. With this information we can hope to learn much about structure formation and the background cosmology. However predictions for cosmological models are usually calculated as distributions over
spatial hypersurfaces. To relate to observation we have to map our light-cone observables into the underlying physical densities, velocities, and other relevant quantities. The aim of this paper is to derive exactly the relation between these quantities, and
provide a fully self-consistent calculation for scalar perturbations in the approximation of linearized general relativity. Relating the physical densities of sources to underlying cosmological parameters is then a separate further problem requiring detailed understanding of the source population and the bias.

This work arose from asking the question: \emph{when comparing observations of the matter power spectrum to theory, in which gauge should the perturbations be calculated?}
For the case of current galaxy surveys, the answer is that it does not really matter since, for observations well within our Hubble volume today, gauge effects are small (we observe wavenumbers with $k/\clh \gg 1$), and cosmic variance is large\footnote{Note that the turnover in the matter power spectrum is determined roughly by the horizon size at matter-radiation equality; differences in gauge are important on scales approaching the horizon scale at the epoch of the observed sources, which is a much larger scale than the turnover for observed redshifts $z\ll 1000$.}.
In practice, what is normally calculated is the synchronous-gauge dark matter power spectrum since this is smooth in $k$ as the Hubble scale is crossed, and then some model is assumed for the bias between source counts and the dark matter density. However, it is well known that redshift distortions are important, which depend on the source velocities, so we are not really directly observing any single source power spectrum but a combination of various effects that contribute to the observed source counts as a function of redshift.

 In this paper we calculate self-consistently (under simple assumptions about bias)
all the various non-stochastic linear effects that enter calculations for the observed angular densities of sources per redshift. The most important of the additional effects have calculated separately before, but for completeness we also include all the velocity and post-Newtonian effects that appear in a full linear analysis. We also discuss the result for sources observed as a function of flux, and give a new result for a magnitude-limited survey including magnification bias, radial-displacement and source evolution effects. In addition we calculate the correlation with the cosmic microwave background temperature, including various velocity and post-Newtonian effects, as well as a new calculation of the cross-correlation with the CMB polarization. Doing a self-consistent analysis means we can be sure that no terms have been prematurely neglected  (within the assumptions).  In Appendix~\ref{sec:linerad} we relate the result for source number counts to that for line radiation.

 Differences between the general-relativistic and Newtonian predictions are in most cases below cosmic variance (depending on exactly what is compared), though potentially a source of bias in cross-correlation studies. For the study of scale-dependent halo bias (as a probe of primordial non-Gaussianity~\cite{Dalal:2007cu})
using a joint analysis of multiple biased tracers, small corrections may be more important since there is only one underlying cosmological perturbation, and hence in principle no cosmic variance on the difference between source counts~\cite{Seljak:2008xr,McDonald:2008sh,White:2008jy}. In particular there is a danger that an incorrect calculation of the large-scale power spectrum could give spurious evidence for scale-dependent bias due to the neglect of various non-negligible effects that are present in the full analysis.

Previous work has examined perturbations to the luminosity distance in detail~\cite{Sasaki87,Pyne:2003bn,Hui:2005nm,Bonvin:2005ps,Barausse:2005nf}. Reference~\cite{Kasai87} considers the total source count to up to a certain redshift in a specific cosmological model, but does not give results for counts as a function of flux or magnitude limit. More recently, while this work was in progress, Refs~\cite{Yoo:2008tj,Yoo:2009au,Yoo:2010ni} have given an analysis more similar to this paper, though our presentation is rather different and our results do not all agree. In particular our numerical results are quantitatively different, in part due to our more consistent treatment of bias.
Our numerical ``CAMB sources'' code for calculating source count angular power spectra and cross-correlations with the CMB and galaxy weak lensing is publicly available\footnote{\url{http://camb.info/sources/}}.

The outline of this paper is as follows. In Sec.~\ref{sec:number} we give a general non-perturbative result for the observed source densities as a function of redshift, then in Sec.~\ref{sec:FRWmisc} we calculate the result for linear scalar perturbations. We use the Newtonian gauge, however since we are calculating an observable the final result is gauge invariant. In Sec.~\ref{sec:luminosity} we include the source luminosities and derive the observed source density for a magnitude-limited survey. In Sec.~\ref{sec:numerical} we discuss the dominant terms in the full result, explain how we perform numerical calculations over a source window function, and present some typical numerical results. We also discuss evolving source populations, and the distinction between the selection function and the underlying background source redshift distribution. Finally in Sec.~\ref{sec:CMB} we give results for the cross-correlation of source counts with the CMB temperature and polarization, including all the various effects that can be important for both low and high redshift sources. In Appendix~\ref{sec:linerad} we relate the result for source number counts to that for line radiation (e.g. 21-cm radiation).

\section{Projected number counts}
\label{sec:number}

We first calculate the observable angular source density count as a function of redshift and angle on the sky in terms of the physical quantities governing the source density, velocity and line-of-sight evolution. Readers not interested in details of the derivation can skip to the main linearized result in Eq.~\eqref{fullcounts_s0} or~\eqref{eq:countsfin_magbias} below. Throughout we use units with $c=1$. Our derivation below, up to and including Eq.~(\ref{eq:counts_conserved}), holds for any metric theory of gravity. However, all numerical results assume general relativity.

We consider a general population of objects with current 4-vector
$J^a = n_s u^a_s$ where $n_s$ is the proper source number density (i.e.\ in the
rest space of the source 4-velocity
$u^a_s$). These are observed in projection by an observer
at $A$ who has 4-velocity $u^a_{oA}$. The observer records $n(\vnhat,z) \ud z \ud\Omega_{oA}$,
the number of sources in direction $\vnhat$ over solid angle $\ud \Omega_{oA}$ with redshift $z$ in a range $\ud z$.
Let $k^a = \ud x^a / \ud \lambda$ be the wave 4-vector of
a light ray on the past lightcone through $A$. If the ray intersects a
source at affine parameter $\lambda$, the redshift of the source is determined from the observed frequency and the emitted frequency in the source rest-frame by
\begin{equation}
1+z = \frac{(k_a u^a_s)|_\lambda}{(k_a u^a_{oA}) |_{\lambda_A}}.
\label{eq:1}
\end{equation}
The infinitesimal angular separation $\delta\theta_J$ of two rays in a bundle at $A$, and the ray-orthogonal connecting vector $\xi_{I}$ between them at affine parameter $\lambda$ along the rays, are related by the Jacobi map $\cld_{IJ}(\lambda)$ (see Refs~\cite{Sachs61,SchneiderBook,Lewis:2006fu}):
\begin{equation}
\xi_{I}(\lambda) = \cld_{IJ}(\lambda) \delta\theta_J,
\end{equation}
where the indices here are components in a 2-dimensional orthonormal basis $\{E^a_I\}$ orthogonal to $k^a$. The Jacobi map is determined by its evolution equation along the ray,
\begin{equation}
\frac{\ud^2\cld_{IJ}}{\ud\lambda^2} = \clt_{IK}\cld_{KJ},
\end{equation}
where the optical tidal matrix is defined in terms of the Riemann tensor by $\clt_{IJ} \equiv - E^b_I E^c_J k^a k^d R_{abcd}$.

We can now use the Jacobi map to relate observed ray angles to
(invariant) areas on the wavefront: a ray bundle with solid angle $\ud \Omega_{oA}$ (as seen by $u^a_{oA}$) has invariant
area $\detD_o \ud \Omega_{oA}$, where $\detD_o$ is the determinant of the
Jacobi map in the $u^a_{oA}$ frame. In incrementing by $\ud \lambda$, the wavefront advances by
a proper distance $\ud \lambda k_a u^a_s$ in the source rest-frame, and so
sweeps up $n_s \detD_o \,k_a u^a_s \ud \lambda\ud \Omega_{oA}$ sources. The general result then follows:
\begin{equation}
n(\vnhat,z) = \detD_o\, k_a J^a \left| \frac{\ud \lambda}{\ud z}\right|,
\label{eq:2}
\end{equation}
where the right-hand side is evaluated at redshift $z$ along the
line of sight with direction $\vnhat$ at the observer.

Under a change of observer, $u^a_{oA} \rightarrow \tilde{u}^a_{oA}$ (so that
$\vnhat \rightarrow \hat{\tilde{\vn}}$ and $z \rightarrow \tilde{z}$),
conservation of the number of sources gives
$\tilde{n}(\hat{\tilde{\vn}},\tilde{z})
\ud \tilde{\Omega}_{oA} \ud \tilde{z} = n(\vnhat,z) \ud \Omega_{oA} \ud z$.
It follows that
\begin{equation}
\tilde{n}(\hat{\tilde{\vn}},\tilde{z}) = \left(\frac{k_a \tilde{u}^a_{oA}}{%
k_a u^a_{oA}}\right) n(\vnhat,z) = \gamma^3 (1+\vnhat \cdot \vv_{\mathrm{rel}})^3
n(\vnhat,z) ,
\label{eq:2b}
\end{equation}
where $\vv_{\mathrm{rel}}$ is the relative velocity of the observers and
$\gamma$ is the associated Lorentz factor.

The result for number counts is related in detail to that for the brightness of
diffuse line radiation from the same sources in Appendix~\ref{sec:linerad};
in summary, the fraction of photons received from a source scales with $1/\detD_o$ (inverse-square law in the background), so the sky brightness is simply $\propto n(\vnhat,z)/\detD_o$.

\section{Projected number counts in flat, almost-FRW models}
\label{sec:FRWmisc}

We aim to evaluate Eq.~(\ref{eq:2}) to linear order
for
a flat, almost-Friedmann-Robertson-Walker (FRW) model with scalar
perturbations. For convenience,
we work in the conformal-Newtonian gauge with metric
\begin{equation}
ds^2 = a^2(\eta) [(1+2\psi)d\eta^2 - (1-2\phi) \delta_{ij} dx^i dx^j].
\label{eq:3}
\end{equation}
We take a zero-shear velocity field $u^a$ along $\partial_\eta$ so that
$u^\mu = a^{-1} (1-\psi) \delta^\mu_0$ and $u_\mu = a(1+\psi)\delta_{\mu 0}$.
This velocity field is the zeroth element of an orthonormal tetrad which
we take to be $(X_0)^a = u^a$ and $X_i \equiv a^{-1} (1+\phi) \partial_i$.

Decomposing the wavevector $k^a = dx^a / d\lambda$ into a direction $e^a$
and frequency $k \cdot u \equiv \epsilon/a$ relative to $u^a$,
we have
\begin{equation}
\frac{d \vx}{d\eta} = (1+\phi + \psi) \ve , \qquad
\frac{d\eta}{d\lambda} = a^{-2} \epsilon (1-\psi),
\label{eq:4}
\end{equation}
where the three-vector $\ve$ comprises the spatial components of
the propagation direction on the spatial triad $X_i$. The geodesic equation
reduces to a simple propagation equation for $\ve$,
\begin{equation}
\frac{d \ve}{d\eta} = - \vgrad_{\perp} (\phi + \psi),
\label{eq:5}
\end{equation}
where $\vgrad_\perp \equiv \vgrad - \ve \ve \cdot \vgrad$, and
an equation for the evolution of the comoving frequency:
\begin{equation}
\frac{\ud \epsilon}{\ud \eta} = - \epsilon
\frac{\ud \psi}{\ud \eta} + \epsilon (\dot{\phi} + \dot{\psi}).
\label{eq:5b}
\end{equation}
Combining
Eqs~(\ref{eq:4}) and (\ref{eq:5}) and solving for the photon
path gives
\begin{equation}
\vx(\vnhat;\eta) = -\ve_A (\eta_A - \eta) + \ve_A \int_{\eta_A}^\eta
(\phi+\psi) \, \ud\eta' - \int_{\eta_A}^\eta (\eta-\eta') \vgrad_\perp
(\phi+\psi)\ud \eta' .
\label{eq:6}
\end{equation}
Note that at the observation point $d\vx / d\eta = (1+\phi+\psi)_A \ve_A$,
so that $-\ve_A$ is the line-of-sight direction for a Newtonian-gauge observer ($u_A^a$). The second
term on the right of Eq.~(\ref{eq:6}) is a radial displacement and
corresponds to the usual (Shapiro) time delay. The third term is the
usual transverse (lensing)
displacement.

We shall need the radial displacement and perturbation to the
conformal time at redshift $z$ along the line of sight. Writing the source
4-velocity as $u^a_s = u^a + v^a$, we have $u^\mu_s = a^{-1} [1-\psi, v^i]$
where $v^i$ are the orthonormal-triad components of $v^a$, and similarly for the observer's 4-velocity $u^a_o = u^a + v^a_{o}$.
It follows that the observed redshift of a source is
\begin{equation}
1+z = \frac{a_A }{a }\frac{\epsilon}{\epsilon_A} \bigl(1+\vnhat \cdot [\vv-\vv_{o A}]\bigr),
\label{eq:7}
\end{equation}
where $\vnhat\equiv -\ve_{oA}$ is the line-of-sight direction on the
observer's triad formed from Lorentz boosting the $(X_0)^a$ and the
$(X_i)^a$. At zeroth order, $\ve_{oA} = \ve_A$.
The ratio of energies follows from integrating Eq.~(\ref{eq:5b}):
\begin{equation}
\frac{\epsilon}{\epsilon_A} = 1 + \psi_A - \psi + \int_{\eta_A}^\eta
(\dot{\phi}+\dot{\psi})\, \ud \eta' ,
\label{eq:8}
\end{equation}
which has the usual Sachs-Wolfe and integrated Sachs-Wolfe (ISW) contributions.
The redshift at $\eta$ along the line of sight is therefore
\begin{equation}
1+z = \frac{a_A}{a(\eta)} \left(1+ \psi_A - \psi + \int_{\eta_A}^\eta
(\dot{\phi} +\dot{\psi})\, \ud \eta' + \vnhat \cdot [\vv-\vv_{oA}] \right) .
\label{eq:9}
\end{equation}
Setting $\eta= \eta_* + \delta \eta$ for a source at observed redshift
$z_*$, where $1+z_* = a_A / a(\eta_*)$, we must have
\begin{equation}
\clh(\eta_*) \delta \eta = \psi_A - \psi + \int_{\eta_A}^{\eta_*}
(\dot{\phi} +\dot{\psi})\, \ud \eta' + \vnhat \cdot [\vv-\vv_{oA}],
\label{eq:10}
\end{equation}
where $\clh$ is the conformal Hubble parameter and the terms on the
right are evaluated on the zero-order lightcone at position
$\vx_A + \vnhat(\eta_A - \eta_*)$ at time $\eta_*$. The radial position
of a photon at observed source redshift $z_*$ then follows from Eq.~(\ref{eq:6}):
\begin{eqnarray}
\chi(\vnhat, z_*) &=& \chi_* + \delta \chi \nonumber \\
		&=& \eta_A-\eta_* - \delta \eta - \int_{\eta_A}^{\eta_*}
(\phi+\psi)\, \ud \eta' .
\label{eq:11}
\end{eqnarray}
To determine the perturbed value of $\detD_o = \detD\ud\Omega_{A}/\ud\Omega_{oA}$, we note that the Jacobi
map is symmetric for linear scalar perturbations. It thus takes the form
\begin{equation}
\cld_{IJ} = \left( \begin{array}{cc} \cld/2 - \gamma_1 & -\gamma_2 \\
				     -\gamma_2 & \cld/2+\gamma_1
		   \end{array}
\right),
\label{eq:12}
\end{equation}
where $\cld$ is the trace of the map and $\gamma_1$ and $\gamma_2$ are the
(first-order) components of the shear. Evaluating the determinant, we
find at linear order that $\det \cld = (\cld/2)^2$. The trace is given in
terms of the convergence $\kappa$ by (see e.g.~Ref.~\cite{Lewis:2006fu})
\begin{equation}
\cld (\vnhat , \eta)/2 = \chi(\vnhat, \eta) a(\eta)[1 - \phi - \kappa(\vnhat,
\eta)],
\label{eq:13}
\end{equation}
where $\chi$ is the perturbed radial position at time $\eta$ and\footnote{Note that other authors (e.g. ~\cite{Bernardeau:2009bm,Yoo:2009au}) have defined the convergence
differently so that it includes additional terms that we are including separately.}
\begin{equation}
2 \kappa(\vnhat,\eta) \equiv - \nabla_{\vnhat}^2
\int_{\eta_A}^{\eta} \frac{(\eta'-\eta)}
{(\eta_A-\eta)(\eta_A - \eta')} (\phi+\psi)\, \ud \eta' .
\label{eq:14}
\end{equation}
The Laplacian here is on the unit sphere.
We actually require $\det \cld$ at \emph{given observed redshift};
this is given by
\begin{equation}
\det \cld(\vnhat,z_*) = a^2(\eta_*) \chi_*^2 \left(1 +
2\frac{\delta\chi}{\chi_*} + 2\clh \delta \eta - 2\phi -2 \kappa \right),
\end{equation}
and hence
\begin{eqnarray}
\detD_o(\vnhat,z_*) &=& \detD(\vnhat,z_*) \frac{\ud\Omega_A}{\ud\Omega_{oA}} \nonumber\\
&=& a^2(\eta_*) \chi_*^2 \left(1 +
2\frac{\delta\chi}{\chi_*} + 2\clh \delta \eta - 2\phi -2 \kappa + 2\vnhat\cdot \vv_{oA}\right).
\end{eqnarray}

We also require the perturbed value of $k_a J^a = n_s k_a u^a_s =
n_s (1+z) \epsilon_A (1+\vnhat\cdot \vv_{oA})/ a_A$ where, recall, $n_s$ is
the proper (physical rather than comoving) number density of sources in their rest frame.
Expanding $n_s$ into a background part,
$\bar{n}_s$, and a perturbed part $\bar{n}_s \delta_n$, and evaluating
the background part at the perturbed time for a source at redshift $z_*$,
we have
\begin{equation}
k_a J^a (\vnhat,z_*) = \frac{(1+z_*)\epsilon_A \bar{n}_s(\eta_*)}{a_A}
\left(1 + \frac{\dot{\bar{n}}_s}{n_s} \delta\eta + \delta_n + \vnhat \cdot
\vv_{oA}\right).
\label{eq:16}
\end{equation}

The final term in Eq.~\eqref{eq:2} is $\ud \lambda / \ud z$.
%
%
We can evaluate this by differentiating Eq.~\eqref{eq:9}:
noting that
\begin{eqnarray}
\ud \eta / \ud \lambda &=& \epsilon (1-\psi)/a^2 \nonumber \\
&=& \frac{\epsilon_A(1+z)}{a a_A}(1-\psi - \vnhat \cdot[ \vv-\vv_{oA}]),
\label{eq:18}
\end{eqnarray}
we find
\begin{eqnarray}
\left| \frac{\ud \lambda}{\ud z} \right|(\vnhat,z_*) &=& \frac{a(\eta_*)a_A}
{\epsilon_A \clh(\eta_*)(1+z_*)^2} \left[1-\left(\frac{\dot{\clh}}{\clh}-\clh
\right)\delta \eta - \frac{1}{\clh}\frac{\ud\psi}{\ud\eta} + \frac{1}{\clh}
(\dot{\phi}+\dot{\psi})
+ \frac{1}{\clh}\vnhat \cdot \frac{\ud \vv}{\ud \eta}
+ \psi + \vnhat\cdot [\vv- \vv_{oA}]\right] .
\label{eq:19}
\end{eqnarray}

Putting these results together, the background result for the number counts per solid angle and redshift is\footnote{This is simply the product of the
comoving source number density ($a^3 \bar{n}_s$) and the differential
comoving volume element per redshift and solid angle.}
\begin{equation}
\bar{n}(z) = (a\chi)^2 \frac{a^2 \bar{n}_s}{a_A \clh}
= \frac{\chi^2}{\clh (1+z)} a^3 \bar{n}_s
,
\label{background}
\end{equation}
where all quantities are evaluated at the source redshift $z$, with fractional perturbation
\begin{multline}
\qquad\Delta_n(\vnhat,z) =
\delta_n + \frac{\dot{\bar{n}}_s}{n_s} \delta\eta
+ 2\frac{\delta\chi}{\chi} + 2\clh \delta \eta - 2\phi -2 \kappa + 3 \vnhat\cdot \vv_{oA}
\\-\left(\frac{\dot{\clh}}{\clh}-\clh
\right)\delta \eta - \frac{1}{\clh}\frac{\ud\psi}{\ud\eta} + \frac{1}{\clh}
(\dot{\phi}+\dot{\psi}) + \frac{1}{\clh}\vnhat \cdot \frac{\ud \vv}{\ud \eta}
+ \psi + \vnhat\cdot [\vv-\vv_{oA}].\qquad
\label{eq:counts_GR}
\end{multline}

For conserved sources  $\dot{\bar{n}}_s/n_s = -3\clh$, so the terms due to constant expansion cancel. More generally, we can write $\Delta_n(\vnhat,z)$ in terms
of the background comoving density ($a^3 \bar{n}_s$) evolution
\begin{multline}
\Delta_n(\vnhat,z) =  \delta_n + \frac{1}{\clh}\vnhat \cdot \frac{\ud \vv}{\ud \eta} + 2\frac{\delta\chi}{\chi}  -2 \kappa+ \frac{\ud \ln (a^3 \bar{n}_s)}{\ud \eta}\delta\eta  \\
+3 \vnhat\cdot \vv_{oA} +
 \vnhat\cdot [\vv-\vv_{oA}]
- \frac{\dot{\clh}}{\clh} \delta \eta - \frac{1}{\clh}\frac{\ud\psi}{\ud\eta}  + \frac{1}{\clh}(\dot{\phi}+\dot{\psi})
  +\psi- 2\phi,
 \label{eq:counts_conserved}
\end{multline}
where terms have been ordered roughly in order of importance on sub-Hubble
scales. This result holds for any metric theory of gravity. However,
if we assume the general-relativistic velocity evolution equation for non-interacting cold particles,
\begin{equation}
\dot{\vv} + \clh \vv + \nabla\psi = 0 ,
\end{equation}
we can simplify further to give
\begin{equation}
\Delta_n(\vnhat,z) =  \delta_n - \frac{1}{\clh}\vnhat \cdot \frac{\partial \vv}{\partial \chi} + 2\frac{\delta\chi}{\chi}  -2 \kappa+ \left[\frac{\ud \ln (a^3 \bar{n}_s)}{\ud \eta} - \frac{\dot{\clh}}{\clh}\right] \delta\eta
 +2 \vnhat\cdot \vv_{oA} + \frac{1}{\clh}\dot{\phi} +\psi- 2\phi.
\label{eq:Deltasimpler}
\end{equation}
The considerable simplification here is from a cancellation between the
change in the Doppler shift across the source volume element due to the time
evolution of $\vnhat \cdot \vv$ and the
change in the Sachs-Wolfe effect due to the radial gradient in $\psi$.

There is a dipole contribution to $\Delta_n(\vnhat,z)$ from the observer's
velocity $\vv_{oA}$. Isolating these terms (including those
implicit in $\delta\eta$ and $\delta \chi$), their contribution can
be written as
\begin{equation}
\Delta^{\vv_{oA}}_n(\vnhat,z) = 3 \vnhat \cdot \vv_{oA} -
\vnhat \cdot \vv_{oA}\frac{1}{\clh} \frac{\ud \ln \bar{n}}{\ud z} .
\end{equation}
This is consistent with the linearized form of Eq.~(\ref{eq:2b}): the
first term on the right is from $(1+\vnhat \cdot \vv_{\mathrm{rel.}})^3$
while the second term is from correcting the observed redshift to
undo the effect of the observer's velocity when
evaluating the background source
counts $\bar{n}(z)$.
Assuming the non-kinematic CMB dipole is $\clo(10^{-5})$, the local peculiar velocity $\vv_{oA}$ can be determined to $\clo(10^{-5})$ from measurements of the observed  (kinematic plus primordial) CMB dipole. If we boost to this frame the $\vnhat\cdot\vv_{oA}$ terms then vanish.

\begin{figure}
\begin{center}
\epsfig{figure=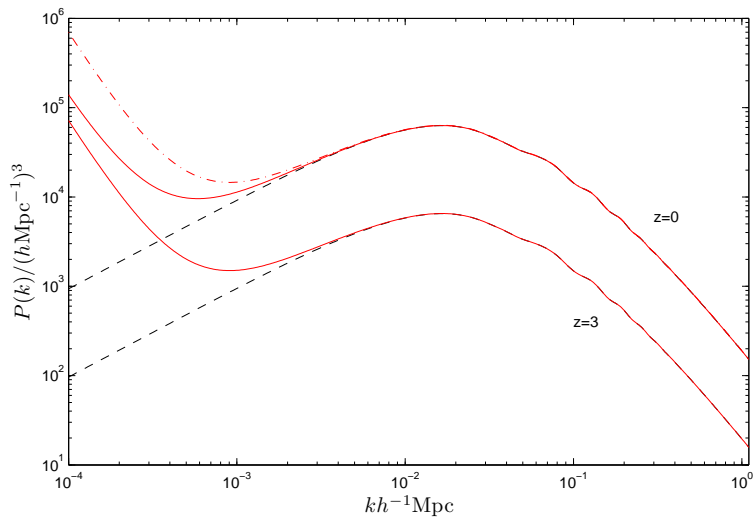,width=10cm}
\caption{
Three-dimensional power spectrum of the Newtonian-gauge counts source $\delta_n$ (solid) and the synchronous gauge source $b \delta_m^\synch$ (dashed) at
$z=0$ and $z=3$ with bias $b=1.5$ (relative to the matter perturbation $\delta_m^\synch$) assuming constant comoving source density. The dash-dotted line shows the spectrum for $\delta_n$ re-scaled (incorrectly) from $z=3$ to $z=0$ with the
scale-independent growth factor $D(\eta)$ appropriate to $\delta_m^\synch$.
On large scales, $\delta_n$ grows more slowly in time than $\delta_m^\synch$
due to
the velocity term in Eq.~(\ref{biaseq}) which goes as $\clh \dot{D}$.
Note that neither power spectrum is directly observable, but the difference between them illustrates the importance of using a full analysis on scales $k\ll 0.01\Mpc^{-1}$.
\label{Pk}}
\end{center}
\end{figure}

Expanding Eq.~(\ref{eq:Deltasimpler}) and dropping local (monopole and
dipole) terms the result is
\begin{multline}
\Delta_n(\vnhat,z) =  \delta_n - \frac{1}{\clh}\vnhat \cdot \frac{\partial \vv}{\partial \chi}  -2 \kappa
+ \left[-\frac{2}{\clh\chi} + \frac{\ud \ln (a^3 \bar{n}_s)}{\clh\ud \eta} - \frac{\dot{\clh}}{\clh^2}\right]
\left[  - \psi - \int^{\eta_A} (\dot{\phi} + \dot{\psi}) \ud \eta + \vnhat\cdot \vv \right]
\\
+ \frac{2}{\chi}\int^{\eta_A}(\phi + \psi)\ud\eta
 + \frac{1}{\clh}\dot{\phi} +\psi- 2\phi.
\label{fullcounts_s0}
\end{multline}
The terms in this equation have a simple interpretation. The first is the
perturbation to the source number density. The second is the
usual redshift-space distortion due to the effect of source velocities on
the differential volume element. The convergence term arises from the
transverse differential effect of gravitational lensing on the volume element
while the following two
terms are from evaluating the zero-order result, Eq.~(\ref{background}),
at the perturbed position and time appropriate to the observed redshift (including the
radial shift due to lensing). The remaining potential terms are from
the perturbation to the spatial metric ($\psi$) in converting $\ud z$
to a proper time interval at source, the differential of that part
of the ISW effect
across the source volume ($\dot \phi$) that is not cancelled by velocity
evolution (see above), and the effect of the spatial metric at source ($2\phi$)
on transverse distances there.

The source number perturbation $\delta_n$ is the Newtonian-gauge quantity. In the simplest case we might expect the synchronous-gauge number perturbation $\delta_n^\synch$
to be related to the matter density perturbation via a constant bias $b$, $\delta_n^\synch = b\delta^\synch_m$, since it is the comoving perturbation (synchronous gauge when perturbations are dominated by pressure-free matter)  that enters directly into the Poisson equation~\cite{Hwang:2006iw,Wands:2009ex}. In this case the Newtonian-gauge number density perturbation is, in Fourier
space\footnote{Our Fourier conventions follow Ref.~\cite{Lewis:2007kz}.},
\begin{equation}
\delta_n= b \delta_m^\synch + \frac{\dot{\bar{n}}_s}{\bar{n}_s}\frac{v}{k}
=
b \delta_m^\synch + \left[ \frac{ \ud \ln (a^3 \bar{n}_s)}{\ud\eta}  - 3\clh\right]\frac{v}{k},
\label{biaseq}
\end{equation}
where $v$ is the Newtonian-gauge velocity of the sources, assumed to follow the matter velocity (no velocity bias). The difference between the Newtonian- and synchronous-gauge matter power spectra is shown in Fig.~\ref{Pk}, and illustrates that it is important to make the distinction on scales approaching the horizon scale.

Our prescription for the bias differs from that in the papers by Yoo et al.~\cite{Yoo:2010ni,Yoo:2009au}, where the assumption is made that the number density of sources $n_s=F(\rho_m)$ is some function of the local matter density. At the spacetime point corresponding to the perturbed position of a source observed at redshift $z$, they linearise this relation about the background matter density at redshift $z$ to define a linear bias relation. This singles out the zero-redshift-perturbation gauge and gives the local relation $\delta_n = b \delta_m$ \emph{in that gauge}. However, it seems clear that the bias is physically due to the local physics of structure formation, and nothing to do with how we observe it: hence in as much that bias is a reasonable model at all, its definition cannot depend on the zero-redshift-perturbation observational gauge. Note further that the synchronous gauge density is also what is calculated in Newtonian collapse and $N$-body simulations~\cite{Hwang:2005xt,Hwang:2006iw,Chisari:2011iq}. For a recent discussion of bias in the general-relativistic context, see~\cite{Baldauf:2011bh}.


The fractional perturbation in the counts is not directly observable
since we cannot observe $\bar{n}(z)$.
Rather, we measure anisotropy in the counts via
$n(\vnhat,z)/\langle n(\vnhat,z) \rangle_{\vnhat}
-1 = \Delta_n(\vnhat,z) - \langle \Delta_n(\vnhat,z) \rangle_{\vnhat}$, where
$\langle \rangle_{\vnhat}$ denotes the angular average over the survey area.
With full-sky coverage, the $l>0$ multipoles of the observed anisotropy
equal those of $\Delta_n(\vnhat,z)$. More generally, the difference in the
observed fractional counts between two directions equals the difference
in $\Delta_n(\vnhat,z)$. Note also that the $l > 1$ moments of
$\Delta_n(\vnhat,z)$ are independent of the observer's velocity in linear theory.

\section{Luminosities}
\label{sec:luminosity}

The linear-theory results derived in the previous section can also be applied to calculate observables other than total source number counts. Of particular interest here is the luminosity distance:  how it affects the flux received from each source, and the predictions for the observed angular densities of a magnitude-limited source sample.

If a source with 4-velocity $u^a_s$
is known to radiate an energy $\ud E = L_s\ud\tau_s$ isotropically
in its rest frame in proper time $\ud\tau_s$ in the form of photons with energy $(k_a u^a_s)|_\lambda$, the number emitted within solid angle $\ud\Omega_s$ is  $\ud N = L_s\ud\tau_s\ud\Omega_s/(4\pi (k_a u^a_s)|_\lambda)$.
These are collected by an observer at $A$, with 4-velocity $u^a_{oA}$ and
equipped with a detector of collecting area $\ud A_A$, in time
$\ud\tau_{oA}$ with energy $(k_a u^a_{oA})|_\lambda$.
Products of areas and solid angles are related
by reciprocity, $(1+z)^2\ud\Omega_s \ud A_s = \ud\Omega_{oA} \ud A_A$ (see e.g. Ref.~\cite{SchneiderBook}), and the Jacobi map relates areas and angles via $\ud A_s = \det \cld_o \ud\Omega_{oA}$. The energy received at $A$ is then given by
\begin{equation}
\ud E_{oA}  = (k_a u^a)|_{\lambda_A} \ud N = \frac{L_s \ud\tau_{oA} \ud A_A}{4\pi(1+z)^4\det\cld_o},
\end{equation}
where we have used $\ud \tau_{oA}/\ud \tau_s = 1+z$.
If there are well-understood sources with known $L_s$ we can therefore measure the luminosity distance, given by
\begin{equation}
d_L(\vnhat,z) = (1+z)^2 [\det \cld_o]^{1/2} ,
\end{equation}
for each source. Our results for the linear-theory Jacobi map at given observed redshift could then be used straightforwardly to work out the observed luminosity distance in a perturbed universe; see  Refs~\cite{Sasaki87,Pyne:2003bn,Hui:2005nm,Bonvin:2005ps,Barausse:2005nf}.

The source number counts can also be measured as a function of source flux, $F=\ud E_{oA}/(\ud A_A\ud \tau_{oA})$, so an observer $u^a_{oA}$ sees $n(\vnhat,z, \ln F)\ud\Omega_{oA}\ud z$ sources per logarithmic flux interval $\ud \ln F$. If the current of sources per log luminosity $L_s$ is $J^a(\ln L_s) = u_s^a n_s(\vx, \eta, \ln L_s)$, we have the observed density
\begin{equation}
n(\vnhat,z,\ln F) = \detD_o\, k_a J^a(\ln L_s) \left| \frac{\ud \lambda}{\ud z}\right|,
\end{equation}
where
\begin{equation}
L_s = 4\pi F (1+z)^4 \detD_o .
\end{equation}

For populations that have smooth distributions in source luminosity, the linear perturbed result is then the same as before (Eq.~\ref{eq:counts_GR}), with some extra terms depending on the slope of the source luminosity function:
\begin{eqnarray}
\Delta_n(\vnhat,z,\ln F) &=&\nonumber \Delta_n(\vnhat,z)|_{\ln \bar{L}_s} + \left.
\frac{\partial \ln \bar{n}_s}{\partial \ln L_s}\right|_{\ln \bar{L}_s} \left( \frac{\detD_o}{a^2(\eta_*)\chi_*^2}-1\right)
\\
&=& \Delta_n(\vnhat,z)|_{\ln \bar{L}_s} + \left.
\frac{\partial \ln \bar{n}_s}{\partial \ln L_s}\right|_{\ln \bar{L}_s} \left( 2\frac{\delta\chi}{\chi_*} + 2\clh \delta \eta - 2\phi -2 \kappa  + 2\vnhat\cdot\vv_{oA}\right),
\end{eqnarray}
where $\Delta_n(\vnhat,z)|_{\ln \bar{L}_s}$ is evaluated for sources with luminosity $\bar{L}_s = 4 \pi F (1+z)^4 a^2(\eta_*)\chi_*^2$, i.e.\ the
luminosity corresponding to flux $F$ in the background.

Apparent magnitudes are defined with conventional factors so that $m = -2.5\log_{10} F+\text{const}$. For a magnitude-limited survey observing all $N(\vnhat,z,m<m_*)\ud z \ud \Omega_{oA}$ sources with magnitudes $m<m_*$ over redshift interval $\ud z$ and in solid angle $\ud \Omega_{oA}$,
the full result for the fractional perturbation to $N$,
neglecting only local terms, is
\begin{multline}
\Delta_N(\vnhat,z,m<m_*) =  \delta_N(L > \bar{L}_{s*})
 - \frac{1}{\clh}\vnhat \cdot \frac{\partial \vv}{\partial \chi}
+ (5s-2) \left[\kappa - \frac{1}{\chi}\int^{\eta_A}(\phi + \psi)\ud\eta  \right] \\
+ \left[\frac{2-5s}{\clh\chi} + 5s- \frac{\partial\ln [a^3 \bar{N}(L > \bar{L}_{s*})]}{\clh\partial \eta} + \frac{\dot{\clh}}{\clh^2}\right]
\left[   \psi + \int^{\eta_A} (\dot{\phi} + \dot{\psi}) \ud \eta - \vnhat\cdot \vv \right]
+ \frac{1}{\clh}\dot{\phi} +\psi + (5s-2)\phi .
 \label{eq:countsfin_magbias}
\end{multline}
Here, $\bar{N}(\eta,L>\bar{L}_{s*})$ is the background number density of
sources with luminosity exceeding $\bar{L}_{s*}$ and its fractional
perturbation is denoted $\delta_N(\vx,\eta,L > \bar{L}_{s*})$. The
quantity
\begin{equation}
s(z,m_*)\equiv \frac{\partial \log_{10} \bar{N}(z,m<m_*)}{\partial m_*} = \frac{\bar{n}_s(\eta,\ln \bar{L}_{s*})}{2.5 \bar{N}(\eta,L > \bar{L}_{s*})} ,
\end{equation}
where the terms in the final expression are evaluated at the background
value of $\eta$ corresponding to $z$.
Note that we have assumed no line-of-sight scattering.

\section{Numerical and approximate results}
\label{sec:numerical}
For non-evolving source populations, Eq.~(\ref{eq:countsfin_magbias}) can be
approximated on small scales as
\begin{eqnarray}
\Delta_N(\vnhat,z,m<m_*) &\approx& \delta_N  -\frac{1}{\clh} \vnhat\cdot\frac{\partial\vv}{\partial\chi}
- \left(\kappa + \frac{\vnhat\cdotp[\vv-\vv_{oA}]}{\clh\chi}\right)(2 - 5s).
\label{deltaN_approx}
\end{eqnarray}
The first term is the obvious underlying perturbation and the second the usual redshift distortion due to velocity gradients~\cite{kaiser87,Hamilton:1997zq}. The third term has two contributions from lensing: the first from the effect of lensing magnification of the background sources~\cite{Gunn67}, and the second slope-dependent correction is the well-known magnification bias~\cite{Matsubara:2000pr,LoVerde:2007ke,Hui:2007cu,Schmidt:2008mb}.  The first of the two contributions proportional to $1/(\clh\chi)$ comes from the
change in the number of background sources per observed solid angle as radial distances are changed~\cite{kaiser87,Szalay:1997cc,Matsubara:1999du,Papai:2008bd,Raccanelli:2010hk}. The second slope-dependent term is the effect of source-dimming due to apparent radial displacements caused by source velocities~\cite{Sasaki87}: a source moving towards us is further away than it appears from its redshift and hence has to be more luminous to give the same observed flux. These $\clo(\vv/\clh\chi)$ terms may be non-negligible compared to the neglected terms if the sources are close, $\chi \ll \clh^{-1}$ (roughly $z\ll 1$).

For numerical work results can be Fourier transformed and expanded into multipoles in $k$ and $l$ space. For example, Eq.~\eqref{deltaN_approx} gives
\begin{equation}
\Delta_{N,l}(k, z) \approx \delta_N j_l(k\chi) + \frac{k v}{\clh} j_l{}''(k\chi) + (2-5s)\left[\frac{v}{\clh\chi}j_l'(k\chi) +  \frac{l(l+1)}{2} \int_0^\chi \ud \chi' \frac{\chi-\chi'}{\chi\chi'} [\phi(\chi') +\psi(\chi')] j_l(k\chi')\right]
\label{eq:jls}
\end{equation}
for $l>1$.
The full result can be calculated similarly, but for brevity we do not quote it here. Integrating over an observed redshift window function, $W(z)$, gives a total fractional perturbation in that window
\begin{equation}
\Delta_{N,l}^W(k) = \int_0^\infty \ud z W(z) \Delta_{N,l}(k,z) =
\int_{0}^{\eta_A}\!\ud\eta
W(\eta) \Delta_{N,l}(k,z_\eta),
\end{equation}
where $z_\eta$ is the redshift at conformal time $\eta$ in the background
and $W(\eta) = (1+z)\clh W(z)$.
 Note that here $W(z)$ is the normalized total distribution of counts in the window, rather than a probability distribution for observing sources; if the number of sources and selection function were constant then $W(z) \propto \bar{n}(z)= a^4\chi^2\bar{n}_s/\clh|_z \propto a\chi^2/\clh|_z $. Defining
 \begin{equation}
 W_{\delta\eta}(\eta) \equiv \left[\frac{2-5s}{\clh\chi} + 5s- \frac{\partial \ln [a^3 \bar{N}(L>\bar{L}_{s*})]}{\clh\partial \eta} + \frac{\dot{\clh}}{\clh^2}\right]_\eta W(\eta),
 \end{equation}
and switching some integration orders, the (full) window-integrated counts transfer function can be calculated using
 \begin{multline}
\Delta^W_{N,l}(k) =
\int_{0}^{\eta_A}\!\ud\eta \biggl[
W(\eta)\left(\delta_N j_l(k\chi) + \frac{k v}{\clh} j_l{}''(k\chi)\right)
+  W_{\delta\eta}(\eta)\left[ \psi j_l(k\chi) + v j_l'(k\chi)   \right]
  +(\dot{\psi} + \dot{\phi})j_l(k\chi)\int_0^\eta W_{\delta\eta}(\eta')\ud\eta' \\
+(\phi+\psi)j_l(k\chi)\left( \int_0^\eta (2-5s) \frac{W(\eta')}{\chi'}\ud\eta'
+\frac{l(l+1)}{2} \int_0^\eta \frac{\chi'-\chi}{\chi\chi'}(2-5s)W(\eta')\ud\eta' \right) \\
+  W(\eta)j_l(k\chi) \left(\frac{1}{\clh}\dot{\phi} +\psi + (5s-2)\phi\right)
\biggr],
\end{multline}
where $\chi' = \eta_A - \eta'$ (and similarly for $\chi$).
 We can further integrate by parts to obtain an integral of a source against $j_l(k\chi)$ (assuming $W(z)$ goes to zero at both ends).
 The linear-theory angular power spectrum can then be calculated using standard line-of-sight Boltzmann codes, giving angular power spectra
\begin{equation}
C_l^{WW'} = 4\pi\int \ud \ln k \clp_{\clr}(k) \Delta_{N,l}^W(k) \Delta_{N,l}^{W'}(k)
\end{equation}
for the cross-correlation between counts in windows $W(z)$ and $W'(z)$.
Here $\clp_{\clr}(k)$ is the dimensionless power spectrum of the primordial
curvature perturbation and the transfer function $\Delta^W_{N,l}(k)$ is for
unit initial curvature perturbation. Cross correlations with the CMB, weak lensing, 21-cm  or other sources can be calculated similarly. Corrections for non-linear evolution can be accounted for approximately by an appropriate re-scaling inside the $k$ integral~\cite{Challinor:2005jy}.

\begin{figure}
\begin{center}
\epsfig{figure=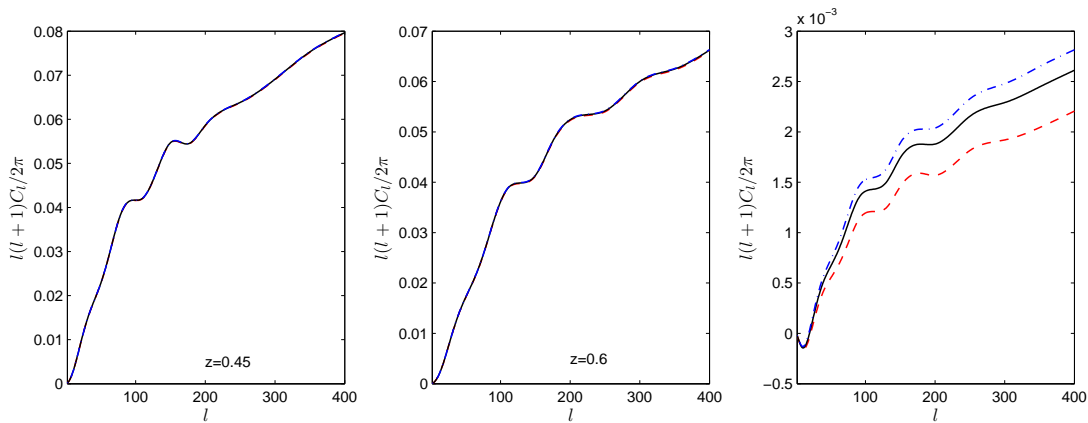,width=14.5cm}
\caption{Angular power spectra of total number counts, based on the approximation of Eq.~(\ref{deltaN_approx}), for Gaussian window
functions $W(z)$ at $z=0.45$ (left; $\sigma_z=0.03$ and constant bias $b=1.5$) and $z=0.6$ (centre; $\sigma_z = 0.05$ and $b=1.95$), and their cross-correlation (right) using non-linear corrections from Halofit~\cite{Smith:2002dz}. Solid lines have no lensing ($2-5s=0$), dashed lines have lensing but no magnification bias ($s=0$), and dot-dashed lines have $s=0.6$. Note that these are barely distinguishable in the auto-power spectra. The window functions are similar to those actually measured using LRG surveys~\cite{Blake:2006kv,Padmanabhan:2006ku}. Lensing is a significant source of correlation when the cross-correlation is otherwise small, c.f.\ Ref.~\cite{LoVerde:2007ke}. The effect is well below cosmic variance on an individual $l$, but above cosmic variance over a range $\Delta_l \sim 100$. For $s=0$ the effect of magnification is negative since a magnified area has less sources per solid angle.
\label{cross}}
\end{center}
\end{figure}

\begin{figure}
\begin{center}
\epsfig{figure=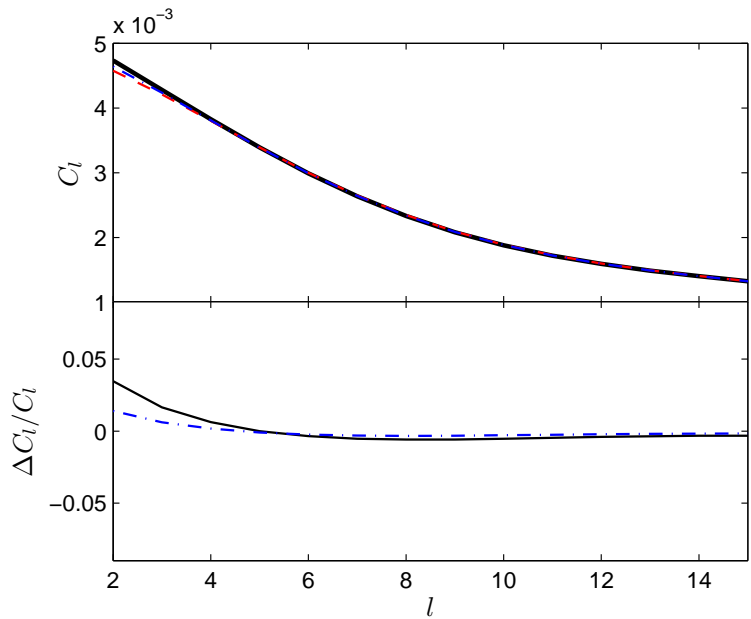,width=10cm}
\caption{Angular power spectra for all sources in a Gaussian window function at $z=0.1$ ($\sigma_z = 0.01$, $b=1$), with (thick solid) and without (dashed) the $\clo(\vnhat\cdot\vv/\clh\chi)$ radial-displacement term. The dot-dashed line shows the equivalent result for a magnitude-limited survey with constant $2-5s = 1$, and the bottom panel shows the fractional differences compared to the result with no radial-displacement terms.
\label{angles}}
\end{center}
\end{figure}

We first consider the effects of the magnification terms in the approximate
result, Eq.~(\ref{deltaN_approx}). These are generally small, but do increase the correlation of counts on different redshift slices. Figure~\ref{cross} shows that the effect can be significant for widely separated slices on small scales when the correlation is otherwise small. As discussed by Ref.~\cite{LoVerde:2006cj} the magnification also has a small effect on the cross-correlation with the CMB (see later discussion in Sec.~\ref{sec:CMB}).
For sources with bias $b$ the relative importance of the magnification terms depends on $(2 - 5s)/b$. Examples for various surveys are given in Ref.~\cite{Hui:2007cu}, and
an example of the large-scale effect is shown in Fig.~\ref{angles} when $2-5s=1$.

Now consider the effect of
the $2\delta\chi/\chi \approx -2\vnhat\cdot[\vv-\vv_{oA}] /(\clh\chi)$ radial-displacement term. This is generally small for distant sources, but can be more important when the sources at are low redshift, as discussed in Refs~\cite{Szalay:1997cc,Matsubara:1999du,Papai:2008bd}. The effect on the large scale power spectrum is shown in Fig.~\ref{angles}.
Having a magnitude limit with small positive $s$ partly cancels the low-redshift radial-displacement effect: a radial displacement gives a larger volume per solid angle, but this is compensated by the fact that the sources appear dimmer because they are further away, so less sources are seen. The effect on the power spectrum for single redshift slices is well below cosmic variance, but may be important for a full redshift survey. The effect of radial displacements on the all-sources correlation function is discussed in Refs~\cite{Matsubara:1999du,Papai:2008bd}.
If there are other selection effects in addition to the magnitude, 
for example on galaxy size~\cite{Schmidt:2009rh}, orientation~\cite{Hirata:2009qz} 
or dust extinction along the line of sight~\cite{Fang:2011hc}, they should also be included.
 Compared to the radial displacement, the effect of the other
$\vnhat\cdot [\vv-\vv_{oA}]$ terms on the power spectrum is down by a factor of  $\clo(\clh\chi)^{-2} \sim 60$ at redshift $z=0.15$ (assuming the source population is not rapidly evolving). At $z\agt 1$, the $\vnhat\cdot [\vv-\vv_{oA}]$ terms are of
equivalent size and all small.

\begin{figure}
\includegraphics[width=10cm]{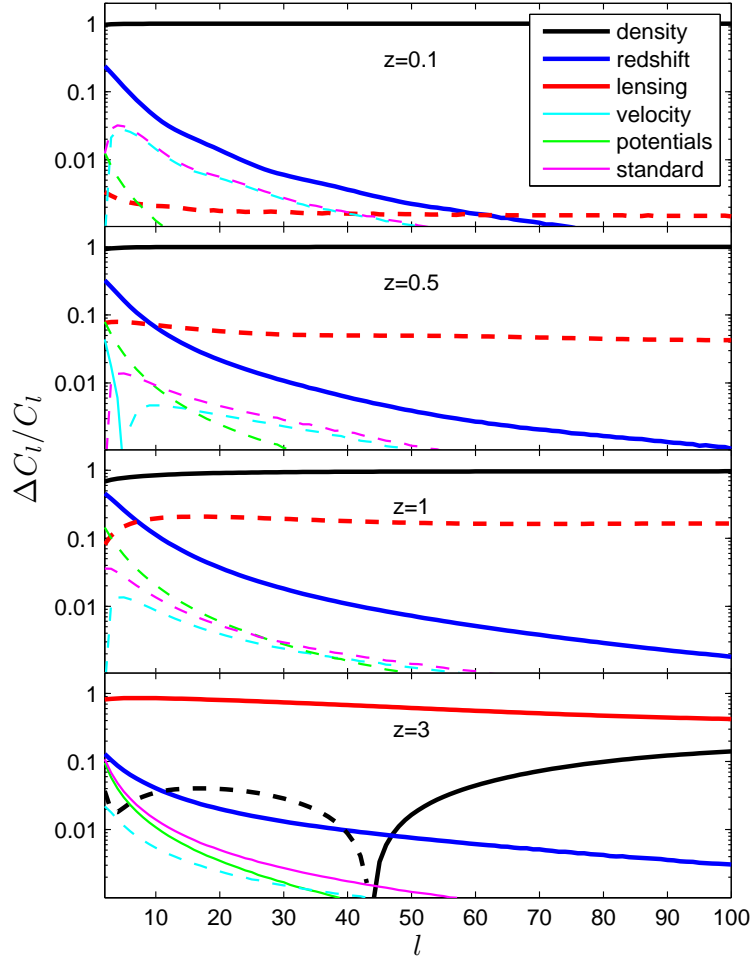}
\caption{
Fractional error compared to the full result for the counts angular power spectrum $C_l$ with broad Gaussian window functions peaking at various redshifts with $\sigma_z=0.3 z$. The error shown is that obtained when various types of term in the full result of Eq.~(\ref{eq:countsfin_magbias}) are individually
neglected. 
The `redshift' term is for redshift-distortions (radial derivative of velocity),
`velocity' terms are proportional to $\vnhat\cdot \vv$ and include the radial displacement effect, `lensing' is the convergence term, and `potentials' includes the effects of gravitational potentials at the source, time delay and the ISW. Solid and dashed lines indicate terms whose neglect reduces and increases the power spectrum respectively.
The `standard' result is the approximation given by Eq.~\eqref{standard_delta} and for this the fractional error compared to the full result is
shown (with solid/dashed lines denoting an excess/deficit).
There is no source evolution and (unrealistically) $b=1$ and  $s=0$.
}
\label{auto_contributions}
\end{figure}


Figure~\ref{auto_contributions} shows the fractional errors that can be made if various terms are neglected when calculating angular power spectra for relatively broad redshift window functions. General-relativistic potential terms are only significant at very low multipoles for high redshift (i.e.\ where the contribution of large-scale modes approaching the Hubble scale can be important). Radial displacement terms are important at low redshift, and lensing (being cumulative) at moderate and high redshift. Although the post-Newtonian effects are generally small compared to cosmic variance, the additional terms entering the observed counts of biased density tracers is independent of bias. If multiple populations tracing the same density field with different bias are being used in a joint analysis the cosmic variance only enters into one eigenvalue of the sample covariance~\cite{Seljak:2008xr,McDonald:2008sh,White:2008jy}, and hence small
theory errors can potentially be rather more important.

\subsection{Total counts distribution}
\label{sec:totalcounts}

We can calculate the angular power spectrum for any combination of redshift window functions $W(z)$. In general these windows are not directly
related to the underlying physical source distribution $\bar{n}(z)$, either because we have chosen to divide up the data into different redshift bins, or because of observational (selection) issues. However we can also consider the special case where we observe all of the sources, and hence wish
to calculate the total angular counts power spectrum over the entire distribution, $W(z) \propto \bar{n}(z) = a^4\chi^2\bar{n}_s/\clh|_z$.

Assuming $s=0$, the velocity terms in Eq.~\eqref{fullcounts_s0} can be written
\begin{eqnarray}
\Delta^v_n(\vnhat,z) &=&   -\frac{1}{\clh}\vnhat \cdot \frac{\partial \vv}{\partial \chi}
   +  \frac{\ud \ln [\chi^2 a^3 \bar{n}_s/\clh]}{\ud \eta} \frac{\vnhat \cdot\vv}{\clh}
\nonumber \\
&=&  -\frac{1}{\clh}\vnhat \cdot \frac{\partial \vv}{\partial \chi}
+ \frac{\ud \ln [ (1+z) \bar{n}(z)]}{\ud\eta}
\frac{\vnhat \cdot\vv}{\clh} .
   \label{vterms}
\end{eqnarray}
Noting that $\ud \vv/ \ud \eta = \dot{\vv} - \partial \vv / \partial \chi$ and
$\ud z = - (1+z) \clh \ud\eta$ in the background,
the perturbation to the velocity terms in the
total counts become
\begin{equation}
\int_0^\infty \ud z \, \bar{n}(z) \Delta^v_n(\vnhat,z) =
\int_0^{\eta_A} \ud \eta\, \left( \frac{\ud}{\ud \eta} [(1+z)\bar{n}(z)
\vnhat\cdot \vv]- (1+z) \bar{n}(z) \vnhat\cdot \dot{\vv}\right) .
\end{equation}
Integrating the total derivative and noting that the boundary terms vanish, we find
\begin{equation}
\int_0^\infty \ud z \, \bar{n}(z) \Delta^v_n(\vnhat,z) =
-\int_0^{\eta_A} \ud \eta\, (1+z) \bar{n}(z) \vnhat\cdot \dot{\vv}.
\end{equation}
%
Since all the sources are being observed, the angular number density is not affected by
changes in the apparent radial distance of each source; the only remaining contributions come from time evolution between the light cone and the perturbed positions.
The result is generally small: a fractional perturbation of $\clo(v)$ rather than $\clo(kv/\clh)$ from the redshift-distortion term alone.
This also means that when calculating the distribution
for all sources it is important to include not only the $\partial\vv/\partial\chi$ term, but all the source velocity terms
since these nearly cancel.
Alternatively in many cases the velocity terms can be neglected entirely, though near cancellation of the velocity terms is a useful numerical consistency check. Similar comments apply to the potential contributions to the
change in redshift distance. However, on the largest scales it is still important to use the correct Newtonian gauge density source, $\delta_n$,
rather than approximating it as proportional to the synchronous-gauge density perturbation.

\begin{figure}
\begin{center}
\epsfig{figure=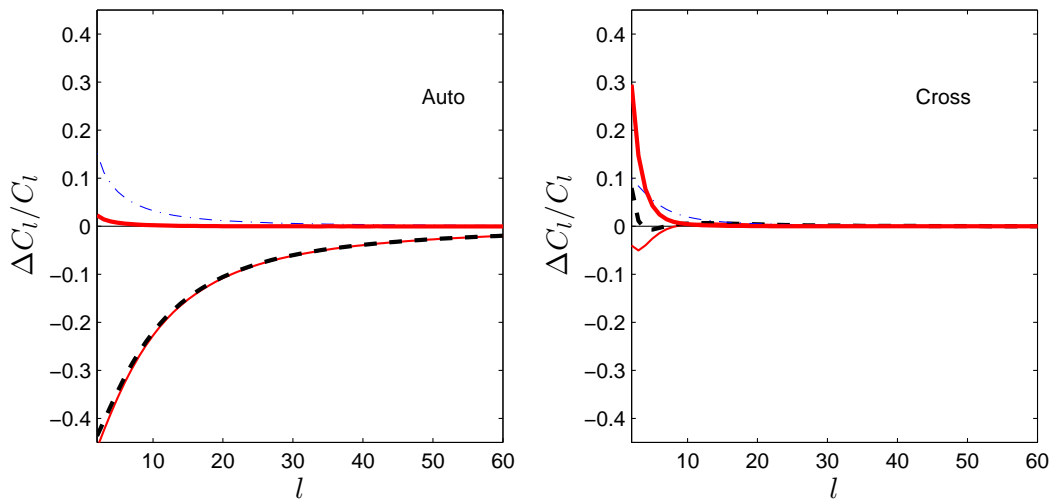,width=14cm}
\caption{
The fractional difference in the counts angular power spectrum (left) and CMB temperature cross-correlation (right), compared with the `standard' result calculated using Eq.~\eqref{standard_delta} for the distribution given in Eq.~\eqref{sampleP}. Thick lines are our limiting full results for the case of no source evolution (solid), in which case the source selection function is non-trivial, and the case when all sources are observed (dashed), so that $P(z)=\bar{n}(z)$ implying source evolution.
The thin red line shows the change in the standard result from dropping the redshift-distortion term; this modification to the `standard' result brings it close
to the full result when all sources are observed as expected from the arguments in Sec.~\ref{sec:totalcounts}.
The blue dashed-dot line shows the change in the standard result when $\delta_N^\synch$ is replaced with the Newtonian gauge $\delta_N$ (no source evolution), which is significantly larger than the difference between the standard and full results.
\label{allcounts}}
\end{center}
\end{figure}

\subsection{Selection function or count distribution?}

Given some observed angle-averaged count distribution $P(z)$, the predicted angular power spectrum for the sample depends on the interpretation of $P(z)$: if $P(z)=\bar{n}(z)$ (i.e. we are observing all sources), then we are calculating the total counts distribution; on the other hand the shape of $P(z)$ may be mainly due to the observational selection function, in which case the underlying total physical count distribution $\bar{n}(z)$ then needs to be specified separately if the source-evolution terms in Eqs~(\ref{fullcounts_s0}) and~(\ref{eq:countsfin_magbias}) are to be included correctly. In general there will be both an observational selection function and some change to the source populations with redshift, and hence $P(z)$ alone does not fully specify the problem.

As an example, we follow Ref.~\cite{Yoo:2009au} by considering a source distribution for a photometric quasar sample,
\be
P(z) \propto z^\alpha \exp\left[-\left(\frac{z}{z_0}\right)^\beta\right]
\label{sampleP}
\ee
with $(\alpha,\beta,z_0) = (3,13, 3.358)$ and hence mean and the peak redshifts
of 2.66 and 3.0 respectively. We take constant $b=2$ and $2-5s=-0.1$.
We compare the full numerical results for the extreme cases of  $P(z)=\bar{n}(z)\propto W(z)$
 (i.e.\ observe all sources) and the limit in which the source population is not evolving ($a^3 \bar{n}_s = \text{const.}$, so $P(z)=\bar{n}(z)p(z)$ where $p(z)$ is the probability of including a given source at redshift $z$ and, from Eq.~\eqref{background}, $\bar{n}(z)\propto a\chi^2/\clh|_z$). For reference we compare to the `standard' result (with no source evolution)
\be
\Delta_{\rm{std}}\equiv \delta_N^\synch - \frac{1}{\clh}\vnhat \cdot \frac{\partial \vv}{\partial \chi} - (2-5s)\kappa = b \delta_m^\synch -  \frac{1}{\clh}\vnhat \cdot \frac{\partial \vv}{\partial \chi} - (2-5s)\kappa.
\label{standard_delta}
\ee
Our results, presented in Fig.~\ref{allcounts},
can be compared with Fig.~2 of Ref.~\cite{Yoo:2009au}, and show a significantly smaller difference
between the full result and the `standard' result for the equivalent cases. The main difference is our alternative treatment of bias (Eq.~\ref{biaseq}) and the explicit treatment (or not) of source evolution terms (c.f. Refs~\cite{Yoo:2008tj,Yoo:2009au}). Nonetheless, we find that the difference between the full and standard result is still
 several percent on the largest scales, and larger for the CMB cross-correlation where there are additional contributions from correlations between count velocity terms and the CMB reionization Doppler signal, and large-mode early-time Sachs-Wolfe contributions (see Sec.~\ref{sec:CMB}). It is also clearly important that the redshift distortion/source evolution terms are incorporated in a consistent manner or much larger differences can be obtained.




\section{CMB cross-correlation}
\label{sec:CMB}

So far we have focused on the correlation function of the counts with themselves. Another useful cosmological probe is the correlation with the cosmic microwave background, primarily due to redshifting of photons as they propagate through the evolving potentials correlated to the source densities (the integrated Sachs-Wolfe effect; ISW)~\cite{Crittenden:1995ak}. There are additional contributions to the correlation due to Doppler terms when CMB photons scatter during reionization~\cite{Giannantonio:2007za}, lensing magnification~\cite{LoVerde:2006cj}, and also non-linear effects such as the Rees-Sciama effect in non-linearly evolving potentials, inhomogeneous reionization and the Sunyaev-Zel'dovich effect. We focus on the linear effects here, so that the fractional CMB temperature anisotropy sourced well after recombination is
\begin{equation}
\Delta_T(\vnhat) \approx \int^{\eta_A} \!\ud\eta \,e^{-\tau}\left(\dot{\tau} \vnhat \cdot \vv + \dot{\psi} + \dot{\phi}\right),
\end{equation}
where $\tau=\tau(\eta)$ is the optical to scattering of CMB photons between $\eta_A$ and $\eta$, and $-\dot{\tau}e^{-\tau}$ is the visibility. This expression neglects small linear contributions from large-scale perturbations at recombination and quadrupole scattering that can easily be included in a full numerical linear calculation (as we do here). Cross-correlation with Eq.~\eqref{eq:countsfin_magbias} gives a variety of terms, dominated by the cross-correlation of the source density and the CMB ISW, but in general with additional non-negligible correlations of the velocities, densities, magnification and potentials. The most important contributions have been calculated separately before, but doing a consistent linear analysis ensures that no relevant effects are missed. Additional velocity and magnification terms are relatively most important when the dominant signal is small, i.e. for sources at high redshift (when the universe is close to matter-dominated) so that the direct ISW contribution from that redshift is small.

\begin{figure}
\begin{center}
\epsfig{figure=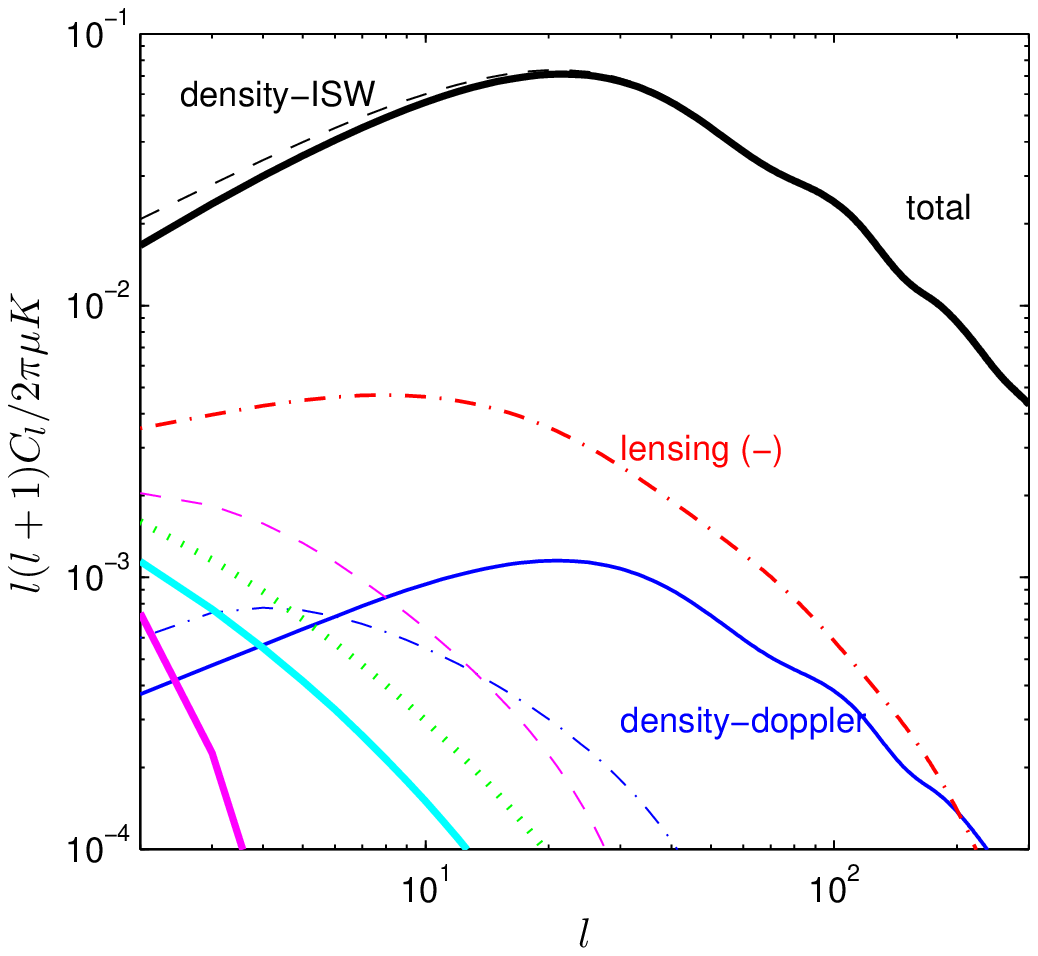,width=8cm}
\epsfig{figure=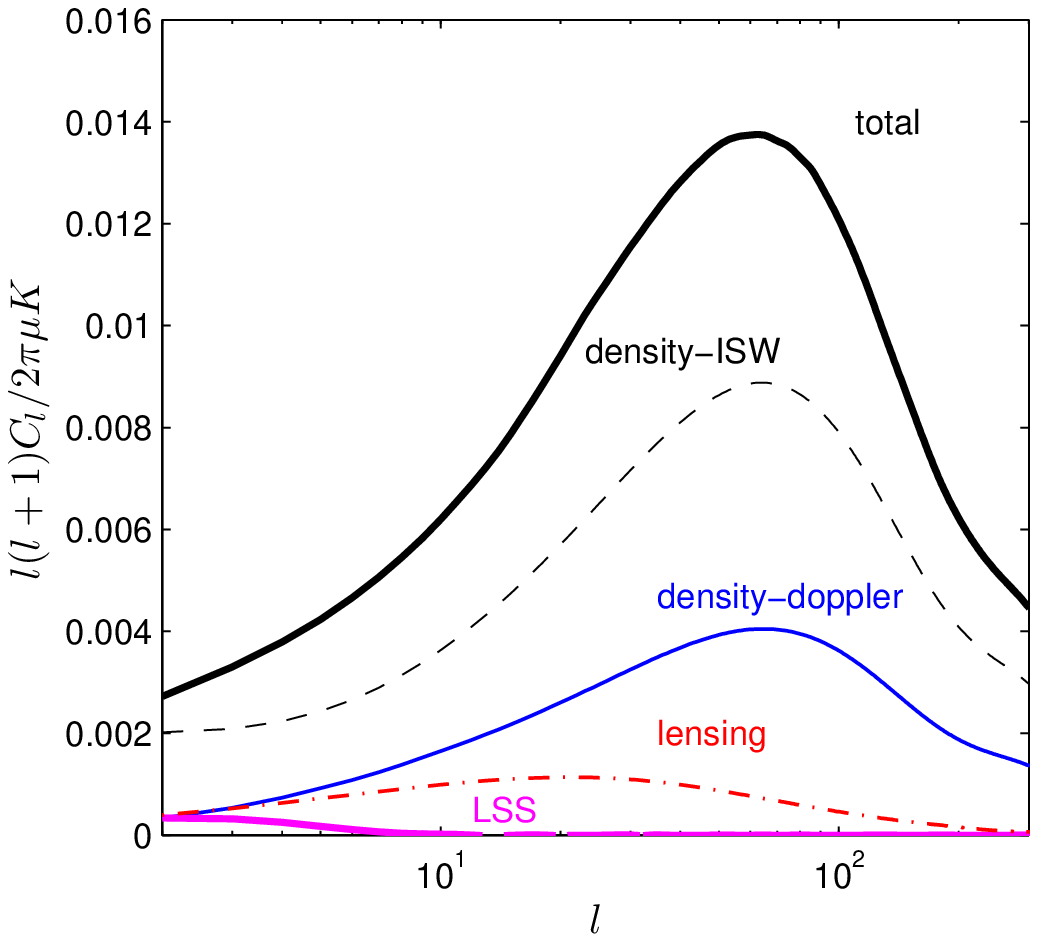,width=8cm}
\caption{Cross-correlation of the CMB temperature and number counts at $z=0.6, \sigma_z=0.05, b=1, s=0$ (left) and $z=3, \sigma_z=0.2, b=2, s=0.42$ (right) assuming no source evolution. Contributions are density-ISW (dashed black), magnification-ISW (dot-dashed red; absolute value), density-Doppler (solid blue), radial velocity gradient-ISW (dashed magenta; redshift distortions), time delay-ISW (thick cyan), total counts-CMB Sachs-Wolfe (thick magenta marked LSS; the CMB contribution is from recombination),  and total (thick solid black). The left figure shows the absolute value of the contributions on a logarithmic scale, and additionally shows the contribution from terms involving non-integrated gravitational potentials (dotted green) and velocity-ISW (lower dot-dashed blue).
The lensing contribution with $s=0$ is negative; the contribution at $z=3$ is low here (right) because $5s-2 = 0.1$.
\label{ISW}}
\end{center}
\end{figure}

\begin{figure}
\begin{center}
\epsfig{figure=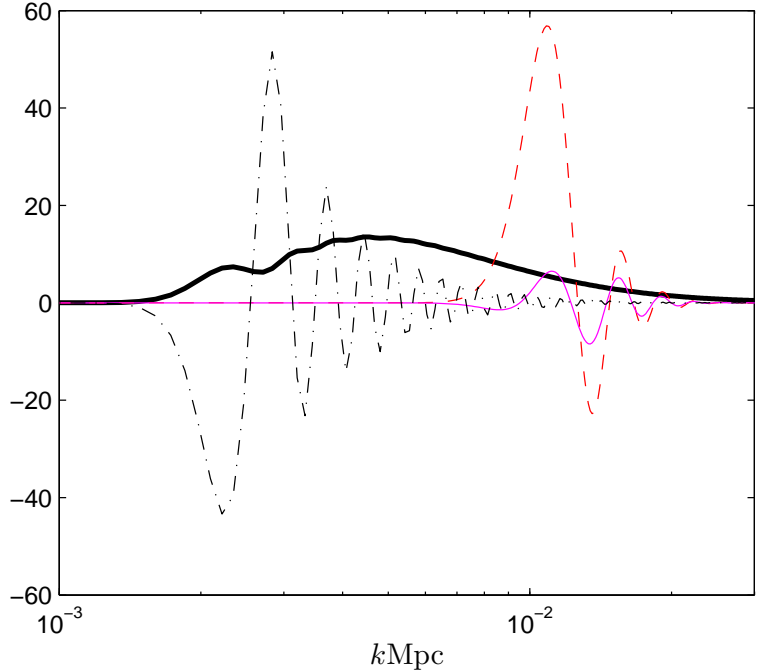,width=10cm}
\caption{Transfer functions at $l=20$ for the ISW (thick solid black) and Doppler (dot-dashed black) contributions to the CMB temperature, and density (thin dashed red) and redshift-distortion (thin solid magenta) contributions to the counts source for a window function peaked at $z=0.6$ with $\sigma_z=0.05$. The corresponding contributions to the $C_l$ are the integrals of the products over $\ln k$ against the primordial power spectrum which is nearly flat. The Doppler and redshift-distortion contributions mostly cancel. CMB transfer functions are scaled by $5\times 10^5$.
\label{ISWtrans}}
\end{center}
\end{figure}

Contributions to the cross-correlation are shown in Fig.~\ref{ISW}. At high redshift, where the potentials are nearly constant, the intrinsic correlation is small and the signal is dominated by lensing unless $2-5s$ is small in which case magnification bias partly cancels the change in angular density of sources from convergence.
At high redshift the electron density is also higher, giving a significant contribution from the CMB Doppler source~\cite{Giannantonio:2007za,LoVerde:2006cj}.  At lower redshifts the correlation of the density and ISW dominates as expected. In addition to the late-time Doppler and ISW sources, there is also a correlation with the Sachs-Wolfe signal from recombination due to very large Hubble-scale modes; these correlate both with the density and velocity count sources on very large scales at the $\alt 10\%$ level (falling rapidly from $l=2$).

On all scales the redshift-distortion and other velocity contributions to the count part of the correlation are small, contrary to the conclusion of Ref.~\cite{Rassat:2009jv}.  Redshift distortions would increase the apparent density of sources over the peak of a matter overdensity, but lead to redshift-space underdensities in the tails even though the potentials contributing to the ISW are still large and the same sign there. Since the redshift-space density averages to zero, the overall correlation is close to zero.
 Mathematically, for a given $l$ and sharp window at $\chi=\chi^*$, the oscillating $j_l''(k\chi^*)$ redshift-distortion term in Eq.~\eqref{eq:jls} is integrated over $\ln k$ against a smooth source for the ISW [$\propto \int \ud\eta(\dot{\phi}+\dot{\psi}) j_l(k\chi)$], giving close to zero; see Fig.~\ref{ISWtrans}. However for numerical work the small few-percent correction on very large scales can easily be included.

\begin{figure}
\begin{center}
\epsfig{figure=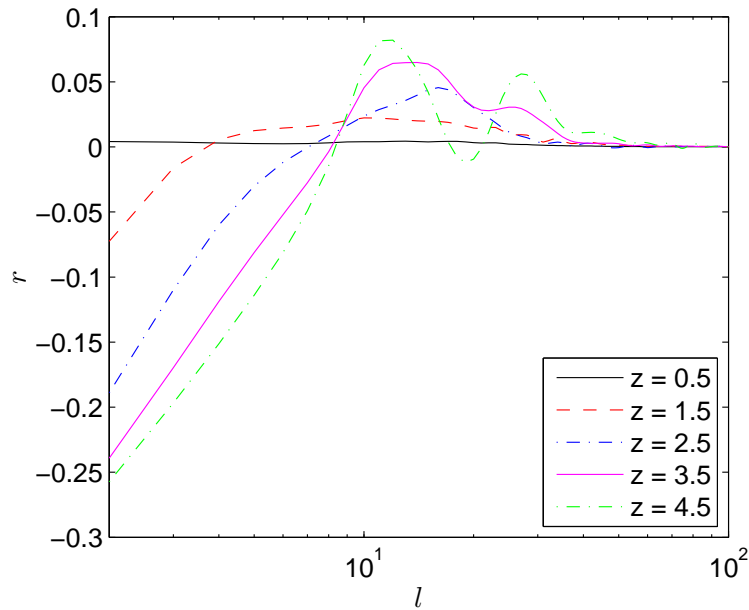,width=10cm}
\caption{The correlation coefficient $r\equiv C_l^{E \Delta}/\sqrt{C_l^{EE} C_l^{\Delta\Delta}}$ between counts ($\Delta$) and CMB $E$-mode polarization for various redshifts, assuming Gaussian redshift bins of width $\sigma_z=0.5$, unit bias, and reionization redshift $z_{\rm re}=11$.
\label{Ecorr}}
\end{center}
\end{figure}

In addition to the correlation between the CMB temperature and counts, there is also some large-scale correlation  with the $E$-mode polarization generated by scattering at reionization. The quadrupole seen by an electron at reionization is generated by the Sachs-Wolfe effect on the electron's last scattering surface; since the Sachs-Wolfe effect is proportional to the gravitational potential and the horizon size at reionization is large (and hence the correlation length corresponds to $l\sim 6$), the polarization generated by Thomson scattering is correlated with the large-scale potentials at significantly lower redshift. For further discussion in the context of lensing see Ref.~\cite{Lewis:2011fk}. For counts at $z\alt 3$ the correlation is in principle marginally detectable on the full sky at $\sim 2\sigma$, but falls rapidly on small scales and low redshift as shown in Fig.~\ref{Ecorr}. However the correlation is significantly larger than calculated in Ref.~\cite{Cooray:2005yj} where only the much smaller low-redshift signal from re-scattering of the ISW signal was included. Combining counts at $z\alt 3$ with CMB lensing to reconstruct the higher redshift potentials, the correlation with polarization is in principle detectable at over $6\sigma$, and hence should be accounted for in any self-consistent full joint analysis.

\section{Conclusions}

We have shown exactly how observable angular source densities as a function of redshift relate to the underlying physical densities, velocities and potentials. In linear theory we recovered various well-known terms, as well as new linear terms that are negligible for modes well inside the Hubble radius.
The gauge-invariant observable results include in a consistent way the different terms that can be described in terms of gravitational lensing, source evolution, radial displacement and other velocity and general-relativistic effects. Our numerical code is available\footnote{\url{http://camb.info/sources/}}, allowing rapid calculation of the auto and cross-spectra for multiple radial window functions, with cosmic shear, and with the CMB and the CMB lensing potential.

We have not addressed in detail the more difficult question of how the number of sources relates to the underlying cosmological perturbations and background cosmology. Nor have we included non-linear effects. Several of these are in fact much more important than the small extra velocity and post-Newtonian terms that we derive here. Apart from non-linear evolution effects on the distribution of the source numbers (and velocities) themselves, there is also lensing of the perturbations~\cite{Schmidt:2008mb,Dodelson:2008qc} and non-linearities in the redshift to real-space mapping~\cite{Scoccimarro:2004tg,Shaw:2008yg}, both of which can have a significant effect on the power spectrum and correlation functions on small
scales.

\section*{Acknowledgements}

\noindent  AL thanks Jaiyul Yoo for communication about his work and sharing his code, 
though we were not ultimately able to establish agreement.
When this work was substantially complete, we learned of related work
by Bonvin and Durrer~\cite{Bonvin2011}. We thank them
for sharing their draft paper and for comparison of numerical
results; where there is overlap, their results
agree with ours.
AL was supported by the Science and Technology Facilities Council (grant numbers ST/F002858/1, PP/C001214/2, and ST/I000976/1).
Some of the calculations for this paper were performed on the
COSMOS Consortium supercomputer within the DiRAC Facility jointly
  funded by STFC and the Large Facilities Capital Fund of BIS.

\appendix

\section{Relating number counts and line radiation}
\label{sec:linerad}

Another important observable is the observed brightness from diffuse line emission (or absorption), especially 21-cm radiation from the spin-flip transition in hydrogen~\cite{Furlanetto:2006jb}. A full linear-theory treatment is complicated due to line-of-sight effects and is discussed in detail
Ref.~\cite{Lewis:2007kz}. Here we make the connection with the
differential number counts discussed in the main text.

In their rest frame, and neglecting the finite line width, the sources with number density $n_s$ emit line radiation
isotropically at frequency $E_0$. In proper time $d\tau$ the number of photons
added per volume within frequency $\ud E$ is
\begin{equation}
\hat{\alpha}\, n_s \, \delta(E-E_0) \ud E \,\ud \tau ,
\label{eq:line1}
\end{equation}
which defines $\hat{\alpha}$. We shall assume $\hat{\alpha}$ is independent of spacetime
position, and neglect scattering and absorption after emission. For a more detailed discussion see Ref.~\cite{Lewis:2007kz}.

We consider collecting the line photons
by an observer with 4-velocity $u^a_A$, equipped with a detector
sensitive to photons in an energy range $\ud E$ about $E$ and with
area $\ud A$ that admits photons in solid angle $\ud \Omega$. The number
of photons collected in proper time $\ud t$ for this observer is
\begin{equation}
\ud \cln = f(E) E^2 \ud E \ud \Omega \ud A \ud t,
\label{eq:relate1}
\end{equation}
in terms of the photon distribution function $f$. Emitting material around affine
parameter $\lambda$ on the line of sight is at redshift $z$. It
emits in its rest-frame at energy $E_0$ and this will contribute to
$\ud \cln$ for $E(1+z) = E_0$. The (invariant)
area of the wavefront associated with rays in the bundle $\ud \Omega$
is $\ud \tilde{A} = \det \cld(\lambda) \ud \Omega$ at $\lambda$,
and the collecting area $\ud A$ subtends a solid angle $\ud \tilde{\Omega}$
in the rest-frame of the emitter at $z$. As the ray advances by
$\ud \lambda$, the wavefront sweeps out a volume $\ud\tilde{A} u^a_s
k_a \ud\lambda$ and it adds to the photons that will be detected in $\ud E$
and $\ud t$ by an amount
\begin{equation}
\hat{\alpha} n_s (1+z)\ud E \delta[E(1+z)-E_0] k_a u^a_s \ud \lambda
\ud \tilde{A} \frac{\ud \tilde{\Omega}}{4\pi} \frac{\ud t}{1+z}.
\label{eq:relate2}
\end{equation}
It follows that
\begin{equation}
\ud \cln = \hat{\alpha} \ud t \ud E \int \ud \lambda \,\left(
k_a J^a \delta[E(1+z)-E_0] \ud \tilde{A} \frac{\ud \tilde{\Omega}}{4\pi}
\right) .
\label{eq:relate3}
\end{equation}
The $\delta$-function allows us to evaluate the integral simply. If we then
compare with Eq.~(\ref{eq:relate1}) and use the reciprocity relation,
$\ud \tilde{A} \ud \tilde{\Omega} = \ud A \ud \Omega / (1+z)^2$, we find
\begin{equation}
Ef(E,\vnhat) = \alpha k_a J^a \left| \frac{\ud \lambda}{\ud z}\right|
= \alpha \frac{n(\vnhat,z)}{\det \cld} ,
\label{eq:relate4}
\end{equation}
where $E(1+z)=E_0$ and $\alpha = \hat{\alpha}/(4\pi E_0^2)$ is another constant.
It is straightforward to show that if we use
Eqs~(\ref{eq:16}) and~(\ref{eq:19}) in the right-hand side of
this expression, we recover the result in Ref.~\cite{Lewis:2007kz}
when line-of-sight scattering and self-absorption is neglected.

\label{lastpage}

\begin{thebibliography}{51}
\expandafter\ifx\csname natexlab\endcsname\relax\def\natexlab#1{#1}\fi
\expandafter\ifx\csname bibnamefont\endcsname\relax
  \def\bibnamefont#1{#1}\fi
\expandafter\ifx\csname bibfnamefont\endcsname\relax
  \def\bibfnamefont#1{#1}\fi
\expandafter\ifx\csname citenamefont\endcsname\relax
  \def\citenamefont#1{#1}\fi
\expandafter\ifx\csname url\endcsname\relax
  \def\url#1{\texttt{#1}}\fi
\expandafter\ifx\csname urlprefix\endcsname\relax\def\urlprefix{URL }\fi
\providecommand{\bibinfo}[2]{#2}
\providecommand{\eprint}[2][]{\url{#2}}

\bibitem[{\citenamefont{Dalal et~al.}(2008)\citenamefont{Dalal, Dore, Huterer,
  and Shirokov}}]{Dalal:2007cu}
\bibinfo{author}{\bibfnamefont{N.}~\bibnamefont{Dalal}},
  \bibinfo{author}{\bibfnamefont{O.}~\bibnamefont{Dore}},
  \bibinfo{author}{\bibfnamefont{D.}~\bibnamefont{Huterer}}, \bibnamefont{and}
  \bibinfo{author}{\bibfnamefont{A.}~\bibnamefont{Shirokov}},
  \bibinfo{journal}{Phys. Rev.} \textbf{\bibinfo{volume}{D77}},
  \bibinfo{pages}{123514} (\bibinfo{year}{2008}), \eprint{0710.4560}.

\bibitem[{\citenamefont{Seljak}(2009)}]{Seljak:2008xr}
\bibinfo{author}{\bibfnamefont{U.}~\bibnamefont{Seljak}},
  \bibinfo{journal}{Phys. Rev. Lett.} \textbf{\bibinfo{volume}{102}},
  \bibinfo{pages}{021302} (\bibinfo{year}{2009}), \eprint{0807.1770}.

\bibitem[{\citenamefont{McDonald and Seljak}(2009)}]{McDonald:2008sh}
\bibinfo{author}{\bibfnamefont{P.}~\bibnamefont{McDonald}} \bibnamefont{and}
  \bibinfo{author}{\bibfnamefont{U.}~\bibnamefont{Seljak}},
  \bibinfo{journal}{JCAP} \textbf{\bibinfo{volume}{0910}}, \bibinfo{pages}{007}
  (\bibinfo{year}{2009}), \eprint{0810.0323}.

\bibitem[{\citenamefont{White et~al.}(2008)\citenamefont{White, Song, and
  Percival}}]{White:2008jy}
\bibinfo{author}{\bibfnamefont{M.}~\bibnamefont{White}},
  \bibinfo{author}{\bibfnamefont{Y.-S.} \bibnamefont{Song}}, \bibnamefont{and}
  \bibinfo{author}{\bibfnamefont{W.~J.} \bibnamefont{Percival}},
  \bibinfo{journal}{Mon. Not. Roy. Astron. Soc.}
  \textbf{\bibinfo{volume}{397}}, \bibinfo{pages}{1348} (\bibinfo{year}{2008}),
  \eprint{0810.1518}.

\bibitem[{\citenamefont{{Sasaki}}(1987)}]{Sasaki87}
\bibinfo{author}{\bibfnamefont{M.}~\bibnamefont{{Sasaki}}},
  \bibinfo{journal}{\mnras} \textbf{\bibinfo{volume}{228}},
  \bibinfo{pages}{653} (\bibinfo{year}{1987}).

\bibitem[{\citenamefont{Pyne and Birkinshaw}(2004)}]{Pyne:2003bn}
\bibinfo{author}{\bibfnamefont{T.}~\bibnamefont{Pyne}} \bibnamefont{and}
  \bibinfo{author}{\bibfnamefont{M.}~\bibnamefont{Birkinshaw}},
  \bibinfo{journal}{Mon. Not. Roy. Astron. Soc.}
  \textbf{\bibinfo{volume}{348}}, \bibinfo{pages}{581} (\bibinfo{year}{2004}),
  \eprint{astro-ph/0310841}.

\bibitem[{\citenamefont{Hui and Greene}(2006)}]{Hui:2005nm}
\bibinfo{author}{\bibfnamefont{L.}~\bibnamefont{Hui}} \bibnamefont{and}
  \bibinfo{author}{\bibfnamefont{P.~B.} \bibnamefont{Greene}},
  \bibinfo{journal}{Phys. Rev.} \textbf{\bibinfo{volume}{D73}},
  \bibinfo{pages}{123526} (\bibinfo{year}{2006}), \eprint{astro-ph/0512159}.

\bibitem[{\citenamefont{Bonvin et~al.}(2006)\citenamefont{Bonvin, Durrer, and
  Gasparini}}]{Bonvin:2005ps}
\bibinfo{author}{\bibfnamefont{C.}~\bibnamefont{Bonvin}},
  \bibinfo{author}{\bibfnamefont{R.}~\bibnamefont{Durrer}}, \bibnamefont{and}
  \bibinfo{author}{\bibfnamefont{M.~A.} \bibnamefont{Gasparini}},
  \bibinfo{journal}{Phys. Rev.} \textbf{\bibinfo{volume}{D73}},
  \bibinfo{pages}{023523} (\bibinfo{year}{2006}), \eprint{astro-ph/0511183}.

\bibitem[{\citenamefont{Barausse et~al.}(2005)\citenamefont{Barausse,
  Matarrese, and Riotto}}]{Barausse:2005nf}
\bibinfo{author}{\bibfnamefont{E.}~\bibnamefont{Barausse}},
  \bibinfo{author}{\bibfnamefont{S.}~\bibnamefont{Matarrese}},
  \bibnamefont{and} \bibinfo{author}{\bibfnamefont{A.}~\bibnamefont{Riotto}},
  \bibinfo{journal}{Phys. Rev.} \textbf{\bibinfo{volume}{D71}},
  \bibinfo{pages}{063537} (\bibinfo{year}{2005}), \eprint{astro-ph/0501152}.

\bibitem[{\citenamefont{{Kasai} and {Sasaki}}(1987)}]{Kasai87}
\bibinfo{author}{\bibfnamefont{M.}~\bibnamefont{{Kasai}}} \bibnamefont{and}
  \bibinfo{author}{\bibfnamefont{M.}~\bibnamefont{{Sasaki}}},
  \bibinfo{journal}{Modern Physics Letters A} \textbf{\bibinfo{volume}{2}},
  \bibinfo{pages}{727} (\bibinfo{year}{1987}).

\bibitem[{\citenamefont{Yoo}(2009)}]{Yoo:2008tj}
\bibinfo{author}{\bibfnamefont{J.}~\bibnamefont{Yoo}}, \bibinfo{journal}{Phys.
  Rev.} \textbf{\bibinfo{volume}{D79}}, \bibinfo{pages}{023517}
  (\bibinfo{year}{2009}), \eprint{0808.3138}.

\bibitem[{\citenamefont{Yoo et~al.}(2009)\citenamefont{Yoo, Fitzpatrick, and
  Zaldarriaga}}]{Yoo:2009au}
\bibinfo{author}{\bibfnamefont{J.}~\bibnamefont{Yoo}},
  \bibinfo{author}{\bibfnamefont{A.~L.} \bibnamefont{Fitzpatrick}},
  \bibnamefont{and}
  \bibinfo{author}{\bibfnamefont{M.}~\bibnamefont{Zaldarriaga}},
  \bibinfo{journal}{Phys. Rev.} \textbf{\bibinfo{volume}{D80}},
  \bibinfo{pages}{083514} (\bibinfo{year}{2009}), \eprint{0907.0707}.

\bibitem[{\citenamefont{Yoo}(2010)}]{Yoo:2010ni}
\bibinfo{author}{\bibfnamefont{J.}~\bibnamefont{Yoo}}, \bibinfo{journal}{Phys.
  Rev.} \textbf{\bibinfo{volume}{D82}}, \bibinfo{pages}{083508}
  (\bibinfo{year}{2010}), \eprint{1009.3021}.

\bibitem[{\citenamefont{Sachs}(1961)}]{Sachs61}
\bibinfo{author}{\bibfnamefont{R.}~\bibnamefont{Sachs}},
  \bibinfo{journal}{Proc. Roy. Soc. Lon.} \textbf{\bibinfo{volume}{264}},
  \bibinfo{pages}{309} (\bibinfo{year}{1961}).

\bibitem[{\citenamefont{Schneider et~al.}(1992)\citenamefont{Schneider, Ehlers,
  and Falco}}]{SchneiderBook}
\bibinfo{author}{\bibfnamefont{P.}~\bibnamefont{Schneider}},
  \bibinfo{author}{\bibfnamefont{J.}~\bibnamefont{Ehlers}}, \bibnamefont{and}
  \bibinfo{author}{\bibfnamefont{E.}~\bibnamefont{Falco}},
  \emph{\bibinfo{title}{Gravitational Lenses}} (\bibinfo{publisher}{Springer},
  \bibinfo{year}{1992}), ISBN \bibinfo{isbn}{0387970703}.

\bibitem[{\citenamefont{Lewis and Challinor}(2006)}]{Lewis:2006fu}
\bibinfo{author}{\bibfnamefont{A.}~\bibnamefont{Lewis}} \bibnamefont{and}
  \bibinfo{author}{\bibfnamefont{A.}~\bibnamefont{Challinor}},
  \bibinfo{journal}{Phys. Rept.} \textbf{\bibinfo{volume}{429}},
  \bibinfo{pages}{1} (\bibinfo{year}{2006}), \eprint{astro-ph/0601594}.

\bibitem[{\citenamefont{Bernardeau et~al.}(2010)\citenamefont{Bernardeau,
  Bonvin, and Vernizzi}}]{Bernardeau:2009bm}
\bibinfo{author}{\bibfnamefont{F.}~\bibnamefont{Bernardeau}},
  \bibinfo{author}{\bibfnamefont{C.}~\bibnamefont{Bonvin}}, \bibnamefont{and}
  \bibinfo{author}{\bibfnamefont{F.}~\bibnamefont{Vernizzi}},
  \bibinfo{journal}{Phys. Rev.} \textbf{\bibinfo{volume}{D81}},
  \bibinfo{pages}{083002} (\bibinfo{year}{2010}), \eprint{0911.2244}.

\bibitem[{\citenamefont{Hwang and Noh}(2006{\natexlab{a}})}]{Hwang:2006iw}
\bibinfo{author}{\bibfnamefont{J.-C.} \bibnamefont{Hwang}} \bibnamefont{and}
  \bibinfo{author}{\bibfnamefont{H.}~\bibnamefont{Noh}},
  \bibinfo{journal}{Phys. Rev.} \textbf{\bibinfo{volume}{D73}},
  \bibinfo{pages}{044021} (\bibinfo{year}{2006}{\natexlab{a}}),
  \eprint{astro-ph/0601041}.

\bibitem[{\citenamefont{Wands and Slosar}(2009)}]{Wands:2009ex}
\bibinfo{author}{\bibfnamefont{D.}~\bibnamefont{Wands}} \bibnamefont{and}
  \bibinfo{author}{\bibfnamefont{A.}~\bibnamefont{Slosar}},
  \bibinfo{journal}{Phys. Rev.} \textbf{\bibinfo{volume}{D79}},
  \bibinfo{pages}{123507} (\bibinfo{year}{2009}), \eprint{0902.1084}.

\bibitem[{\citenamefont{Lewis and Challinor}(2007)}]{Lewis:2007kz}
\bibinfo{author}{\bibfnamefont{A.}~\bibnamefont{Lewis}} \bibnamefont{and}
  \bibinfo{author}{\bibfnamefont{A.}~\bibnamefont{Challinor}},
  \bibinfo{journal}{Phys. Rev.} \textbf{\bibinfo{volume}{D76}},
  \bibinfo{pages}{083005} (\bibinfo{year}{2007}), \eprint{astro-ph/0702600}.

\bibitem[{\citenamefont{Hwang and Noh}(2006{\natexlab{b}})}]{Hwang:2005xt}
\bibinfo{author}{\bibfnamefont{J.-C.} \bibnamefont{Hwang}} \bibnamefont{and}
  \bibinfo{author}{\bibfnamefont{H.}~\bibnamefont{Noh}}, \bibinfo{journal}{Gen.
  Rel. Grav.} \textbf{\bibinfo{volume}{38}}, \bibinfo{pages}{703}
  (\bibinfo{year}{2006}{\natexlab{b}}), \eprint{astro-ph/0512636}.

\bibitem[{\citenamefont{Chisari and Zaldarriaga}(2011)}]{Chisari:2011iq}
\bibinfo{author}{\bibfnamefont{N.~E.} \bibnamefont{Chisari}} \bibnamefont{and}
  \bibinfo{author}{\bibfnamefont{M.}~\bibnamefont{Zaldarriaga}}
  (\bibinfo{year}{2011}), \eprint{1101.3555}.

\bibitem[{\citenamefont{Baldauf et~al.}(2011)\citenamefont{Baldauf, Seljak, Senatore, and Zaldarriaga}}]{Baldauf:2011bh}
\bibinfo{author}{\bibfnamefont{T.} \bibnamefont{Baldauf}} \bibnamefont{and}
  \bibinfo{author}{\bibfnamefont{U.}~\bibnamefont{Seljak}} \bibnamefont{and}
\bibinfo{author}{\bibfnamefont{L.} \bibnamefont{Senatore}} \bibnamefont{and}
\bibinfo{author}{\bibfnamefont{M.} \bibnamefont{Zaldarriaga}}
  (\bibinfo{year}{2011}), \eprint{1106.5507}.

\bibitem[{\citenamefont{{Kaiser}}(1987)}]{kaiser87}
\bibinfo{author}{\bibfnamefont{N.}~\bibnamefont{{Kaiser}}},
  \bibinfo{journal}{\mnras} \textbf{\bibinfo{volume}{227}}, \bibinfo{pages}{1}
  (\bibinfo{year}{1987}).

\bibitem[{\citenamefont{{Hamilton}}(1998)}]{Hamilton:1997zq}
\bibinfo{author}{\bibfnamefont{A.~J.~S.} \bibnamefont{{Hamilton}}}, in
  \emph{\bibinfo{booktitle}{The Evolving Universe}}, edited by
  \bibinfo{editor}{\bibfnamefont{D.}~\bibnamefont{{Hamilton}}}
  (\bibinfo{year}{1998}), vol. \bibinfo{volume}{231} of
  \emph{\bibinfo{series}{Astrophysics and Space Science Library}}, pp.
  \bibinfo{pages}{185--+}, \eprint{astro-ph/9708102}.

\bibitem[{\citenamefont{{Gunn}}(1967)}]{Gunn67}
\bibinfo{author}{\bibfnamefont{J.~E.} \bibnamefont{{Gunn}}},
  \bibinfo{journal}{\apj} \textbf{\bibinfo{volume}{147}}, \bibinfo{pages}{61}
  (\bibinfo{year}{1967}).

\bibitem[{\citenamefont{Matsubara}(2000)}]{Matsubara:2000pr}
\bibinfo{author}{\bibfnamefont{T.}~\bibnamefont{Matsubara}},
  \bibinfo{journal}{Astrophys. J. Lett.} \textbf{\bibinfo{volume}{537}},
  \bibinfo{pages}{77} (\bibinfo{year}{2000}), \eprint{astro-ph/0004392}.

\bibitem[{\citenamefont{LoVerde et~al.}(2008)\citenamefont{LoVerde, Hui, and
  Gaztanaga}}]{LoVerde:2007ke}
\bibinfo{author}{\bibfnamefont{M.}~\bibnamefont{LoVerde}},
  \bibinfo{author}{\bibfnamefont{L.}~\bibnamefont{Hui}}, \bibnamefont{and}
  \bibinfo{author}{\bibfnamefont{E.}~\bibnamefont{Gaztanaga}},
  \bibinfo{journal}{Phys. Rev.} \textbf{\bibinfo{volume}{D77}},
  \bibinfo{pages}{023512} (\bibinfo{year}{2008}), \eprint{0708.0031}.

\bibitem[{\citenamefont{Hui et~al.}(2007)\citenamefont{Hui, Gaztanaga, and
  LoVerde}}]{Hui:2007cu}
\bibinfo{author}{\bibfnamefont{L.}~\bibnamefont{Hui}},
  \bibinfo{author}{\bibfnamefont{E.}~\bibnamefont{Gaztanaga}},
  \bibnamefont{and} \bibinfo{author}{\bibfnamefont{M.}~\bibnamefont{LoVerde}},
  \bibinfo{journal}{Phys. Rev.} \textbf{\bibinfo{volume}{D76}},
  \bibinfo{pages}{103502} (\bibinfo{year}{2007}), \eprint{0706.1071}.

\bibitem[{\citenamefont{Schmidt et~al.}(2008)\citenamefont{Schmidt, Vallinotto,
  Sefusatti, and Dodelson}}]{Schmidt:2008mb}
\bibinfo{author}{\bibfnamefont{F.}~\bibnamefont{Schmidt}},
  \bibinfo{author}{\bibfnamefont{A.}~\bibnamefont{Vallinotto}},
  \bibinfo{author}{\bibfnamefont{E.}~\bibnamefont{Sefusatti}},
  \bibnamefont{and} \bibinfo{author}{\bibfnamefont{S.}~\bibnamefont{Dodelson}},
  \bibinfo{journal}{Phys. Rev.} \textbf{\bibinfo{volume}{D78}},
  \bibinfo{pages}{043513} (\bibinfo{year}{2008}), \eprint{0804.0373}.

\bibitem[{\citenamefont{{Szalay} et~al.}(1998)\citenamefont{{Szalay},
  {Matsubara}, and {Landy}}}]{Szalay:1997cc}
\bibinfo{author}{\bibfnamefont{A.~S.} \bibnamefont{{Szalay}}},
  \bibinfo{author}{\bibfnamefont{T.}~\bibnamefont{{Matsubara}}},
  \bibnamefont{and} \bibinfo{author}{\bibfnamefont{S.~D.}
  \bibnamefont{{Landy}}}, \bibinfo{journal}{\apjl}
  \textbf{\bibinfo{volume}{498}}, \bibinfo{pages}{L1+} (\bibinfo{year}{1998}),
  \eprint{arXiv:astro-ph/9712007}.

\bibitem[{\citenamefont{{Matsubara}}(2000)}]{Matsubara:1999du}
\bibinfo{author}{\bibfnamefont{T.}~\bibnamefont{{Matsubara}}},
  \bibinfo{journal}{\apj} \textbf{\bibinfo{volume}{535}}, \bibinfo{pages}{1}
  (\bibinfo{year}{2000}), \eprint{arXiv:astro-ph/9908056}.

\bibitem[{\citenamefont{Papai and Szapudi}(2008)}]{Papai:2008bd}
\bibinfo{author}{\bibfnamefont{P.}~\bibnamefont{Papai}} \bibnamefont{and}
  \bibinfo{author}{\bibfnamefont{I.}~\bibnamefont{Szapudi}},
  \bibinfo{journal}{\mnras} \textbf{\bibinfo{volume}{389}},
  \bibinfo{pages}{292} (\bibinfo{year}{2008}), \eprint{0802.2940}.

\bibitem[{\citenamefont{Raccanelli et~al.}(2010)\citenamefont{Raccanelli,
  Samushia, and Percival}}]{Raccanelli:2010hk}
\bibinfo{author}{\bibfnamefont{A.}~\bibnamefont{Raccanelli}},
  \bibinfo{author}{\bibfnamefont{L.}~\bibnamefont{Samushia}}, \bibnamefont{and}
  \bibinfo{author}{\bibfnamefont{W.~J.} \bibnamefont{Percival}}
  (\bibinfo{year}{2010}), \eprint{1006.1652}.

\bibitem[{\citenamefont{Challinor and Lewis}(2005)}]{Challinor:2005jy}
\bibinfo{author}{\bibfnamefont{A.}~\bibnamefont{Challinor}} \bibnamefont{and}
  \bibinfo{author}{\bibfnamefont{A.}~\bibnamefont{Lewis}},
  \bibinfo{journal}{Phys. Rev.} \textbf{\bibinfo{volume}{D71}},
  \bibinfo{pages}{103010} (\bibinfo{year}{2005}), \eprint{astro-ph/0502425}.

\bibitem[{\citenamefont{Smith et~al.}(2003)}]{Smith:2002dz}
\bibinfo{author}{\bibfnamefont{R.~E.} \bibnamefont{Smith}} \bibnamefont{et~al.}
  (\bibinfo{collaboration}{The Virgo Consortium}), \bibinfo{journal}{Mon. Not.
  Roy. Astron. Soc.} \textbf{\bibinfo{volume}{341}}, \bibinfo{pages}{1311}
  (\bibinfo{year}{2003}), \eprint{astro-ph/0207664}.

\bibitem[{\citenamefont{Blake et~al.}(2007)\citenamefont{Blake, Collister,
  Bridle, and Lahav}}]{Blake:2006kv}
\bibinfo{author}{\bibfnamefont{C.}~\bibnamefont{Blake}},
  \bibinfo{author}{\bibfnamefont{A.}~\bibnamefont{Collister}},
  \bibinfo{author}{\bibfnamefont{S.}~\bibnamefont{Bridle}}, \bibnamefont{and}
  \bibinfo{author}{\bibfnamefont{O.}~\bibnamefont{Lahav}},
  \bibinfo{journal}{\mnras} \textbf{\bibinfo{volume}{374}},
  \bibinfo{pages}{1527} (\bibinfo{year}{2007}), \eprint{astro-ph/0605303}.

\bibitem[{\citenamefont{Padmanabhan et~al.}(2007)}]{Padmanabhan:2006ku}
\bibinfo{author}{\bibfnamefont{N.}~\bibnamefont{Padmanabhan}}
  \bibnamefont{et~al.} (\bibinfo{collaboration}{SDSS}), \bibinfo{journal}{Mon.
  Not. Roy. Astron. Soc.} \textbf{\bibinfo{volume}{378}}, \bibinfo{pages}{852}
  (\bibinfo{year}{2007}), \eprint{astro-ph/0605302}.

\bibitem[{\citenamefont{LoVerde et~al.}(2007)\citenamefont{LoVerde, Hui, and
  Gaztanaga}}]{LoVerde:2006cj}
\bibinfo{author}{\bibfnamefont{M.}~\bibnamefont{LoVerde}},
  \bibinfo{author}{\bibfnamefont{L.}~\bibnamefont{Hui}}, \bibnamefont{and}
  \bibinfo{author}{\bibfnamefont{E.}~\bibnamefont{Gaztanaga}},
  \bibinfo{journal}{Phys. Rev.} \textbf{\bibinfo{volume}{D75}},
  \bibinfo{pages}{043519} (\bibinfo{year}{2007}), \eprint{astro-ph/0611539}.

\bibitem[{\citenamefont{Schmidt et~al.}(2009)\citenamefont{Schmidt, Rozo,
  Dodelson, Hui, and Sheldon}}]{Schmidt:2009rh}
\bibinfo{author}{\bibfnamefont{F.}~\bibnamefont{Schmidt}},
  \bibinfo{author}{\bibfnamefont{E.}~\bibnamefont{Rozo}},
  \bibinfo{author}{\bibfnamefont{S.}~\bibnamefont{Dodelson}},
  \bibinfo{author}{\bibfnamefont{L.}~\bibnamefont{Hui}}, \bibnamefont{and}
  \bibinfo{author}{\bibfnamefont{E.}~\bibnamefont{Sheldon}},
  \bibinfo{journal}{Phys. Rev. Lett.} \textbf{\bibinfo{volume}{103}},
  \bibinfo{pages}{051301} (\bibinfo{year}{2009}), \eprint{0904.4702}.

\bibitem[{\citenamefont{Hirata}(2009)}]{Hirata:2009qz}
\bibinfo{author}{\bibfnamefont{C.~M.} \bibnamefont{Hirata}},
  \bibinfo{journal}{Mon. Not. Roy. Astron. Soc.}
  \textbf{\bibinfo{volume}{399}}, \bibinfo{pages}{1074} (\bibinfo{year}{2009}),
  \eprint{0903.4929}.

\bibitem[{\citenamefont{Fang et~al.}(2011)\citenamefont{Fang, Hui, Menard, May,
  and Scranton}}]{Fang:2011hc}
\bibinfo{author}{\bibfnamefont{W.}~\bibnamefont{Fang}},
  \bibinfo{author}{\bibfnamefont{L.}~\bibnamefont{Hui}},
  \bibinfo{author}{\bibfnamefont{B.}~\bibnamefont{Menard}},
  \bibinfo{author}{\bibfnamefont{M.}~\bibnamefont{May}}, \bibnamefont{and}
  \bibinfo{author}{\bibfnamefont{R.}~\bibnamefont{Scranton}}
  (\bibinfo{year}{2011}), \eprint{1105.3421}.

\bibitem[{\citenamefont{Crittenden and Turok}(1996)}]{Crittenden:1995ak}
\bibinfo{author}{\bibfnamefont{R.~G.} \bibnamefont{Crittenden}}
  \bibnamefont{and} \bibinfo{author}{\bibfnamefont{N.}~\bibnamefont{Turok}},
  \bibinfo{journal}{Phys. Rev. Lett.} \textbf{\bibinfo{volume}{76}},
  \bibinfo{pages}{575} (\bibinfo{year}{1996}), \eprint{astro-ph/9510072}.

\bibitem[{\citenamefont{Giannantonio and
  Crittenden}(2007)}]{Giannantonio:2007za}
\bibinfo{author}{\bibfnamefont{T.}~\bibnamefont{Giannantonio}}
  \bibnamefont{and}
  \bibinfo{author}{\bibfnamefont{R.}~\bibnamefont{Crittenden}},
  \bibinfo{journal}{Mon. Not. Roy. Astron. Soc.}
  \textbf{\bibinfo{volume}{381}}, \bibinfo{pages}{819} (\bibinfo{year}{2007}),
  \eprint{0706.0274}.

\bibitem[{\citenamefont{Rassat}(2009)}]{Rassat:2009jv}
\bibinfo{author}{\bibfnamefont{A.}~\bibnamefont{Rassat}}
  (\bibinfo{year}{2009}), \eprint{0902.1759}.

\bibitem[{\citenamefont{Lewis et~al.}(2011)\citenamefont{Lewis, Challinor, and
  Hanson}}]{Lewis:2011fk}
\bibinfo{author}{\bibfnamefont{A.}~\bibnamefont{Lewis}},
  \bibinfo{author}{\bibfnamefont{A.}~\bibnamefont{Challinor}},
  \bibnamefont{and} \bibinfo{author}{\bibfnamefont{D.}~\bibnamefont{Hanson}},
  \bibinfo{journal}{JCAP} \textbf{\bibinfo{volume}{1103}}, \bibinfo{pages}{018}
  (\bibinfo{year}{2011}), \eprint{1101.2234}.

\bibitem[{\citenamefont{Cooray and Melchiorri}(2006)}]{Cooray:2005yj}
\bibinfo{author}{\bibfnamefont{A.}~\bibnamefont{Cooray}} \bibnamefont{and}
  \bibinfo{author}{\bibfnamefont{A.}~\bibnamefont{Melchiorri}},
  \bibinfo{journal}{JCAP} \textbf{\bibinfo{volume}{0601}}, \bibinfo{pages}{018}
  (\bibinfo{year}{2006}), \eprint{astro-ph/0511054}.

\bibitem[{\citenamefont{Dodelson et~al.}(2008)\citenamefont{Dodelson, Schmidt,
  and Vallinotto}}]{Dodelson:2008qc}
\bibinfo{author}{\bibfnamefont{S.}~\bibnamefont{Dodelson}},
  \bibinfo{author}{\bibfnamefont{F.}~\bibnamefont{Schmidt}}, \bibnamefont{and}
  \bibinfo{author}{\bibfnamefont{A.}~\bibnamefont{Vallinotto}},
  \bibinfo{journal}{Phys. Rev.} \textbf{\bibinfo{volume}{D78}},
  \bibinfo{pages}{043508} (\bibinfo{year}{2008}), \eprint{0806.0331}.

\bibitem[{\citenamefont{Scoccimarro}(2004)}]{Scoccimarro:2004tg}
\bibinfo{author}{\bibfnamefont{R.}~\bibnamefont{Scoccimarro}},
  \bibinfo{journal}{Phys. Rev.} \textbf{\bibinfo{volume}{D70}},
  \bibinfo{pages}{083007} (\bibinfo{year}{2004}), \eprint{astro-ph/0407214}.

\bibitem[{\citenamefont{Shaw and Lewis}(2008)}]{Shaw:2008yg}
\bibinfo{author}{\bibfnamefont{J.~R.} \bibnamefont{Shaw}} \bibnamefont{and}
  \bibinfo{author}{\bibfnamefont{A.}~\bibnamefont{Lewis}},
  \bibinfo{journal}{Phys. Rev.} \textbf{\bibinfo{volume}{D78}},
  \bibinfo{pages}{103512} (\bibinfo{year}{2008}), \eprint{0808.1724}.

\bibitem[{\citenamefont{Bonvin and Durrer}(2011)}]{Bonvin2011}
\bibinfo{author}{\bibfnamefont{C.}~\bibnamefont{Bonvin}} \bibnamefont{and}
  \bibinfo{author}{\bibfnamefont{R.}~\bibnamefont{Durrer}}
  (\bibinfo{year}{2011}), \eprint{1105.5280}

\bibitem[{\citenamefont{Furlanetto et~al.}(2006)\citenamefont{Furlanetto, Oh,
  and Briggs}}]{Furlanetto:2006jb}
\bibinfo{author}{\bibfnamefont{S.}~\bibnamefont{Furlanetto}},
  \bibinfo{author}{\bibfnamefont{S.~P.} \bibnamefont{Oh}}, \bibnamefont{and}
  \bibinfo{author}{\bibfnamefont{F.}~\bibnamefont{Briggs}},
  \bibinfo{journal}{Phys. Rept.} \textbf{\bibinfo{volume}{433}},
  \bibinfo{pages}{181} (\bibinfo{year}{2006}), \eprint{astro-ph/0608032}.

\end{thebibliography}

\providecommand{\aj}{Astron. J. }\providecommand{\apj}{Astrophys. J.
  }\providecommand{\apjl}{Astrophys. J.
  }\providecommand{\mnras}{MNRAS}\providecommand{\aap}{Astron.
  Astrophys.}\providecommand{\aj}{Astron. J. }\providecommand{\apj}{Astrophys.
  J. }\providecommand{\apjl}{Astrophys. J.
  }\providecommand{\mnras}{MNRAS}\providecommand{\aap}{Astron. Astrophys.}

\end{document}